\documentclass{ptptex}
\usepackage{graphicx}

\begin{document}
\def\etal{{\it et al.\ }}
\graphicspath{{fig/}}
\def\percent{{\%}}

\title{Molecular Dynamics for Dense Matter}

\author{Toshiki \textsc{Maruyama}$^1$, 
Gentaro \textsc{Watanabe}$^{2,3,4}$,
and
Satoshi \textsc{Chiba}$^{5,1}$}

\inst{
$^1$Advanced Science Research Center, Japan Atomic Energy Agency, Shirakata Shirane 2-4, Tokai, Ibaraki, 319-1195 Japan\\
$^2$Asia Pacific Center for Theoretical Physics (APCTP), POSTECH, San 31, Hyoja-dong, Nam-gu, Pohang, Gyeongbuk 790-784, Korea\\
$^3$Department of Physics, POSTECH, San 31, Hyoja-dong, Nam-gu, Pohang, Gyeongbuk 790-784, Korea\\
$^4$Nishina Center, RIKEN, 2-1 Hirosawa, Wako, Saitama 351-0198, Japan\\
$^5$Research Laboratory for Nuclear Reactors, Tokyo Institute of Technology, 2-12-1-N1-9 Ookayama, Meguro-ku, Tokyo, 152-8550, Japan

}

\maketitle

\abstract{
We review a molecular dynamics method for nucleon many-body systems called the quantum molecular dynamics (QMD) and our studies using this method.
These studies address the structure and the dynamics of nuclear matter relevant to the neutron star crusts, supernova cores, and heavy-ion collisions.
A key advantage of QMD is that we can study dynamical processes of nucleon many-body systems without any assumptions on the nuclear structure.
First we focus on the inhomogeneous structures of low-density nuclear matter consisting not only of spherical nuclei but  also of nuclear ``pasta'', i.e., rod-like and slab-like nuclei.
%The pasta phases have been predicted by many works with assuming the nuclear shape.
%Here, 
We show that the pasta phases can appear in the ground and equilibrium states of nuclear matter without assuming nuclear shape.
Next we show our simulation of compression of nuclear matter which corresponds to the collapsing stage of supernovae.
With increase of density, a crystalline solid of spherical nuclei  change to a triangular lattice of rods by connecting  neighboring nuclei.
Finally, we discuss the fragment formation in expanding nuclear matter.
Our results suggest that a generally accepted scenario based on the liquid-gas phase transition is not plausible at lower temperatures.
}

% We review a molecular dynamics method for nucleon many-body systems called the quantum molecular dynamics (QMD) and our studies using this method.
% These studies address the structure and the dynamics of nuclear matter relevant to the neutron star crusts, supernova cores, and heavy-ion collisions.
% A key advantage of QMD is that we can study dynamical processes of nucleon many-body systems without any assumptions on the nuclear structure.
% First, we focus on the rod-like and slab-like nuclei, so-called the "pasta" nuclei, in low-density nuclear matter in neutron stars and supernovae.
% Here, we show that the pasta phases can appear in the ground and equilibrium states of nuclear matter without assuming nuclear shape.
% Next, we show our simulation of compression of nuclear matter in collapsing supernova cores.
% In this simulation, we directly observe the dynamics of a transition from a crystalline solid of spherical nuclei to a triangular lattice of rod-like nuclei induced by the compression.
% Finally, we discuss the fragment formation of expanding nuclear matter in heavy-ion collisions.
% Unlike a generally accepted scenario based on the gas-liquid phase transition, our simulation supports another picture in which cold solid-like matter breaks into fragments by the formation of cracks.
% 

\section{Introduction}
Due to the progress of computers, numerical simulations became increasingly capable in tackling complicated problems in nuclear physics.
%Due to the progress of computers, ability and possibility of numerical simulations are growing in nuclear physics.
Generally, numerical simulations can be classified into two types: macroscopic and microscopic simulations.
The former, macroscopic simulations, deal directly with  the macroscopic quantities which we are interested in.
We need to introduce physics models which describe how the quantities are connected with each other. 
On the other hand, microscopic simulations are based on the degrees of freedom of the constituent elements. 
The necessary inputs are equations of motion and the interactions among the elements.
The properties of the total system are obtained later by analyzing the resultant information of these constituent elements.
Microscopic simulations have several advantages: 
1) We need only a few assumptions on the model. 
2) We may obtain unexpected results. 
3) We may find a physical principle (the law governing the elements) if we obtain suitable observables.

\begin{figure}
\begin{center}
\includegraphics[width=0.5\textwidth]{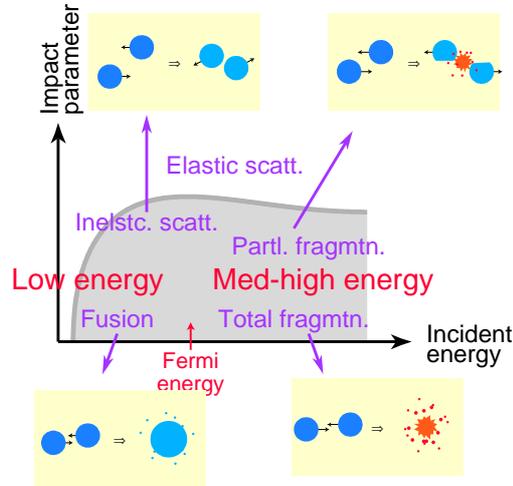}
\caption{\label{reaction-mechanism}
Schematic diagram of heavy-ion reaction mechanism.
}
\end{center}
\end{figure}

This review article is about the molecular dynamics (MD) simulations of nuclear matter.
%In this article, we employ a molecular dynamics (MD) simulation for the study of nuclear matter.
First, we give an overview of the history of the MD models in nuclear physics. 
The simulation study of nuclear dynamics originated from the formulation of the time-dependent Hartree-Fock (TDHF) theory in 1930 \cite{Dirac}.
TDHF deals with the time evolution of many-fermion systems and is an approximation of the time-dependent Schr\"odinger equation with the use of a single Slater determinant.
However, it was in the 1970's that the TDHF was first solved numerically \cite{Bonche}.
Due to the limitation of computer power, only  low-energy phenomena were studied in the early stage.
%The subject was limited to low-energy phenomena.
As the computational power drastically increased after the 1980's, applications to higher-energy phenomena with larger numbers of degrees of freedom and also improvements of the framework to include correlations, etc., have been made.
However, TDHF cannot describe the heavy-ion collision process at higher energies where the degrees of freedom drastically increase.
The reaction mechanism depends on the incident energy and the impact parameter, as shown in Fig.\ \ref{reaction-mechanism}.
Particularly, the Fermi energy is the key quantity to characterize the mechanism.
It is because the reaction mechanism is determined by the competition between Fermi motion inside the nuclei and the relative motion of colliding nuclei.
%Above the Fermi energy (medium-high energy), the overlapped region will break into fragments, i.e., fragmentation occurs.
%
Above the Fermi energy (medium -- high energy), the region where the colliding nuclei overlap with each other will break into fragments, i.e., fragmentation occurs.
This fragmentation process is driven by the energy of the relative motion between the two nuclei converted into thermal excitations, which are generated by two-body collisions. 
Each collision is regarded as a transition from a Slater determinant into another. 
Such a change of the wavefunction cannot be described by TDHF with a single Slater determinant.

At higher energies, the above-mentioned two-particle collision becomes an important process which determines the reaction mechanism in the heavy-ion collisions.
The time evolution of the phase-space distribution function in heavy-ion collisions is described by a Boltzmann-type equation of motion (EOM), with a smooth change by a Newtonian equation and dissipation by the two-body collision process, which is called Boltzmann-Uehling-Uhlenbeck (BUU), Vlasov-Uehling-Uhlenbeck (VUU), or Boltzmann-Nordheim-Vlasov (BNV) equation.
If one omits the collision term, this framework can be regarded as a classical limit ($\hbar\rightarrow 0$) of TDHF \cite{CYWong}, i.e., the Vlasov equation.
This Boltzmann-type equation can be numerically solved by a test-particle method:
The fluid elements in a 6-dimensional phase space are replaced by a classical particle and the phase-space distribution function is obtained by counting the number of those particles in the 6-dimensional mesh (three dimensions for coordinate space and the other three dimensions for momentum space).
For sufficiently large numbers of test particles, the 6-dimensional particle density is conserved throughout the time evolution in a mean-field potential described by the Vlasov equation. 
The two-body collision process of test particles violates the conservation of 6-dimensional phase-space distribution.
%% This violation corresponds to the collision term.
In the 1980's many works on the heavy-ion collisions in the medium -- high energy region have been made via BUU simulations with the test-particle method.

\begin{figure}
\begin{center}
\includegraphics[width=0.5\textwidth]{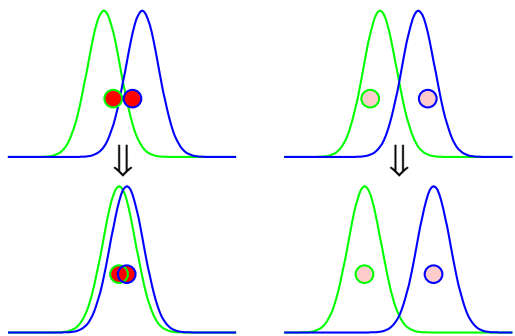}
\caption{Schematic explanation on the time evolution of two-particle correlation.
The curves indicate the distribution functions and the circles are their representative test particles.
Refer to the text for details.
\label{figCorrelation}}
\end{center}
\end{figure}

Molecular dynamics simulation for nuclear systems has been developed in the late 1980's.
Aichelin and St\"ocker have proposed quantum molecular dynamics (QMD) model to simulate heavy-ion collisions from medium to high energies \cite{QMD}.
This framework is obtained by reducing the number of test particles in BUU simulations so that each particle represents one nucleon.
%They have reduced the number of test particles of BUU simulation so that each particle represent one nucleon.
By this reduction of test particles, it became possible to describe many-body correlation of the system.
Let us take the two-body correlation as an example.
If two nucleons stay within a distance so that their distribution functions (solid curves in Fig.\ \ref{figCorrelation}) overlap with each other, the representative two test particles (circles in Fig.\ \ref{figCorrelation}) can be very close to each other (corresponding to the left panel of Fig.\ \ref{figCorrelation}) and also can be far (right panel).
In the former case, those two nucleons may be bound to form a cluster while, in the latter case, they may diverge from each other as time passes.	
However, with a huge number of test particles, we obtain only one time evolution of the distribution function which corresponds to the average of many events. 
This example of the two-body correlation effect on the fragment formation is related to the variation between time evolutions of different events.
It is also natural that the two-body correlation is important for the description of spatial fluctuation in a event.
In the mean-field calculation, on the other hand, both fluctuations in space and the variation between events will be washed out.

%The molecular dynamics model proposed by Aichelin and St\"ocker \cite{refQMD,refQMDreview} is named quantum molecular dynamics (QMD).
The QMD model assumes a direct product of nucleon single-particle wavefunctions as a total wavefunction and a Lagrangian with a non-relativistic kinetic energy and a potential energy from effective interactions among nucleons.
The single-particle wavefunction is assumed to be a Gaussian wavepacket with a fixed width.
The EOM of the wavefunction is derived from a variational principle with the above Lagrangian, and results in a classical EOM with a Hamiltonian as a function of coordinates of those Gaussian wavepackets.
In the QMD model, the stochastic two-body collision process is added to the time evolution by the Hamilton EOM.
The final state of the two-body collision process is checked so that it obeys the Pauli principle, i.e., the condition on the phase-space density.

QMD is named ``quantum'' due to (1) the many-body correlation or fluctuation in density caused by the EOM and the collision term, (2) the stochasticity in the collision process, (3) the Pauli blocking in the final state of collision, and (4) the use of Gaussian wavepackets for single-particle wavefunctions.
However, the actual feature of QMD simulation is rather classical.
First, the time evolution of the system concerns only the centroids of wavepackets.
Their width in the coordinate space, which is a fixed parameter, appears only in the interaction among the particles by means of the double folding.
The width in the momentum space gives rise to a part of the kinetic energy.
However, this energy is spurious, i.e., it will never be effective, since it is constant during the time evolution.
%Second, the fermion many-body nature, e.g., the Fermi motion, is not basically taken into account.
Second, the Pauli principle, which yields the fermionic momentum distribution, is not basically taken into account.
When the number of the test particles per nucleon is large enough in the Boltzmann type simulation, the phase-space density is conserved in the moving frame of a fluid element.
In spite of the many-body nature obtained by the reduction of the number of test particles, it sacrifices the fermionic nature of the system. 

One of the most serious problems of QMD is in the description of the ground state.
Due to the lack of fermionic characteristics, the energy minimum states of QMD model violate the Pauli principle, and all the particles degenerate into zero in the momentum space so that they overestimate the binding energy.
%the energy-minimum states of the many-body system (nuclei) break the Pauli principle, and all the particles degenerate into zero in the momentum space so that they over-estimate the binding energy. and all the particles degenerate into zero in the momentum space so that they over-estimate the binding energy.
We cannot use the energy-minimum states as the initial conditions of collision simulations.
%If we dare to prepare the initial conditions with appropriate binding energies, the constituent nucleons are moving.
If we prepare  initial conditions with appropriate binding energies, the constituent nucleons would be moving.
This motion makes the initial condition unstable against the emission of nucleons.
To take the fermionic characteristics into account, we need to introduce explicitly the antisymmetrization of the wavefunction \cite{refFMD,refAMD,FMDreview}.

Fermionic molecular dynamics (FMD) \cite{refFMD} and antisymmetrized molecular dynamics (AMD) \cite{refAMD} have been proposed in 1990 and 1992, respectively.
They assume a Slater determinant of Gaussian wavepackets as the wavefunction of the system.
In FMD, the widths of nucleons are time-dependent variables and the kinetic energies of wavepackets are not spurious.
In AMD, the widths of wavepackets are constant in time but the zero-point center-of-mass kinetic energies of fragments are removed in a phenomenological way.
They have succeeded in describing the ground state properties of light nuclei as well as the dynamical processes of low-energy heavy-ion collisions.
The problem is, however, a huge amount of computing cost to solve the equations of motion of FMD and AMD, which is proportional to the fourth power of the particle number $N$ (cf. $\propto N^2$ for QMD).
Thus the use of FMD and AMD has been limited to small systems with the total number of particles up to a few hundreds.

%\subsection{Pauli potential in QMD}

In this situation, a new phenomenological way to mimic the Pauli principle was introduced in QMD \cite{Pei91}.
Wilets \etal \cite{Wilets} and then Dorso \etal \cite{Dorso} developed a repulsive two-body potential so-called the Pauli potential. 
It is a function of not only the distance in the coordinate space, but also of the distance in the momentum space.
This repulsive potential acts between nucleons with the same spin and isospin so that it prevents those particles from coming close in the phase space.
Note that, in this framework, simulated ideal Fermi gases contain the potential energy which comes from the Pauli potential.
It is counted as a part of the nuclear potential energy when one determines the parameters of effective potential.
Due to the momentum dependence of the Pauli potential, constituent nucleons have non-zero values of the momentum in the ground state keeping their velocities at zero; thus the above-mentioned spurious emission of nucleons is avoided.

Since the appearance of the QMD model with the Pauli potential, it became possible to carry out simulations of systems with a large number of nucleons.
One interesting target was low-density (below the saturation density) nuclear matter in compact stars such as crusts of neutron stars and cores of supernovae.
%In low-density nuclear matter, exotic structures called ``nuclear pasta'' have been expected.
In low-density nuclear matter, exotic structures called ``nuclear pasta'' have been predicted by Ravenhall \etal \cite{Ravenhall83} and Hashimoto \etal \cite{Hashimoto84}.
There, nuclear matter cannot be uniform due to a negative partial pressure of nucleons and should be clusterized.
With increasing density, the shape of the cluster changes from droplet, rod, slab, tube, bubble, and then uniform.
The name of ``pasta'' comes from the similarity of the rods to ``spaghetti'' and the slabs to ``lasagna'', etc.
Since Ravenhall \etal and Hashimoto \etal have proposed, many works have been done on nuclear matter with pasta structures.
Most of them are based on the Wigner-Seitz (WS) approximation, 
%% which assumes a geometrical symmetry with the dimensionality of 1, 2, and 3.
% In other words, 
  in which a unit cell with the dimensionality 1, 2, and 3 is replaced by the same volume of the plate, cylinder, and sphere, respectively.
The WS approximation is useful and saves much CPU time.
However, the use of WS cell should be a strong constraint on the structure and only simple structures are allowed.
On the other hand, MD simulation is a microscopic framework which does not need any assumption on the structure and the reaction mechanism.
%Since the QMD model has been developed capable of simulating systems with a large number of particles, we have applied it to the low-density nuclear matter and its inhomogeneous structures.
Since a QMD model is capable of simulating systems with a huge number of particles, we have applied it to low-density nuclear matter and its inhomogeneous structures.
In Sec.\ \ref{sec-matter}, we present the results of our QMD simulations.
%of static nuclear matter and also nuclear matter with a density which changes dynamically.

Apart from adapting QMD to low-energy phenomena, other efforts have been made to the opposite direction, i.e., an attempt to describe high-energy phenomena.
Sorge \etal have proposed relativistic quantum molecular dynamics (RQMD) \cite{refRQMD} in 1989.
% which is based on a fully Lorentz-covariant treatment of many-body systems.
% The Hamiltonian consists of a linear combination of Lorentz-covariant constraints, i.e., those of mass-shell conditions and those which connect the relative times of the particles.
% The former constraints include two-body interactions which depend on Lorentz-invariant center-of-mass (CM) distances.
% %In RQMD, each particles have their own times, i.e., we need to fix their times in order to carry out the simulation.
% In the latter, the time-fixation is made so that two particles have equal times in their CM system.
% In addition to the equations of motion by the Hamiltonian, two-body collisions are taken into account using the CM distances.
The main improvements of RQMD from QMD are (1) the Lorentz covariance in the interaction, kinematics, and the two-body collisions and (2) the inclusion of baryon resonances, strange particles, and the string excitations in the two-body collision process.
Simulations of high-energy heavy-ion collisions have been carried out to analyze experiments with $E/A\approx 1$ -- $200$ GeV at the SIS, AGS, and SPS facilities.
At further high energies, interaction between particles becomes less important and the production of mesons and excitations of baryon resonances are essentially important.
A set of computational codes called ``ultrarelativistic QMD'' (UrQMD) is developed with a collision term highly tuned-up to include various kinds of baryons, mesons, and their excited states \cite{refUrQMD}.
It is distributed on the internet and is often used by many people for simulations of heavy-ion collisions at RHIC experiments.

\section{Molecular dynamics approach to nuclear matter\label{sec-formulation}}

%\subsection{Formulation}

%------------------------------------------------------------------------
\subsection{The total wavefunction and the equation of motion}
%------------------------------------------------------------------------

In QMD, each nucleon state is represented by a Gaussian wavefunction of width $\lambda$,
\begin{equation}
\phi_i({\bf r}) = \frac{1}{(2\pi \lambda^2)^{3/4}} \exp \left[
                - \frac{({\bf r} - {\bf R}_i)^2}{4\lambda^2} +
                  {i} {\bf r} \cdot {\bf P}_i \right],
\end{equation}
where ${\bf R}_i$ and ${\bf P}_i$ are the centers of position and momentum of $i$ th nucleon, respectively.
The total wavefunction is assumed to be a direct product of these wavefunctions.
Thus the one-body distribution function is obtained by the Wigner transform of the wavefunction,
\begin{equation}
f({\bf r},{\bf p}) =  \sum_i { f_i({\bf r},{\bf p}) },
\label{f0}
\end{equation}
\begin{equation}
f_i({\bf r},{\bf p})  =  8 \cdot
                        \exp\left[-{({\bf r}-{\bf R}_i)^2\over 2\lambda^2}
                      -{2\lambda^2({\bf p}-{\bf P}_i)^2}\right].
\label{fi}
\end{equation}

The equations of motion of ${\bf R}_i$ and ${\bf P}_i$ are given by the Hamiltonian equations
\begin{equation}
\dot{{\bf R}}_i =   \frac{\partial H}{\partial {\bf P}_i},
\;\;\;\;
\dot{{\bf P}}_i = - \frac{\partial H}{\partial {\bf R}_i},
\label{newton00}
\end{equation}
and a stochastic nucleon-nucleon collision term. 
Hamiltonian $H$ consists of the kinetic energy and the energy of the two-body effective interactions.

%------------------------------------------------------------------------
\subsection{Effective interactions}\label{secEffectiveInt}
%------------------------------------------------------------------------

Our Hamiltonian is separated into several parts as follows,
\begin{equation}
H  =  T + V_{\rm Pauli} + V_{\rm local} + V_{\rm MD}\;,
\label{ham0}
\end{equation}
where $T$, $V_{\rm Pauli}$, $V_{\rm local}$, and $V_{\rm MD}$ are the kinetic energy, the Pauli potential, the local (momentum-independent) potential, and the momentum-dependent potential parts, respectively.

The Pauli potential \cite{Pei92,Boa88,Ohn92,Mar95} is introduced to mimic the fermionic properties in a semiclassical way.
This phenomenological potential prohibits nucleons of the same spin $\sigma$ and isospin $\tau$ from coming close to each other in the phase space.
Here we employ the Gaussian form of the Pauli potential \cite{Dorso,Pei92} as
\begin{equation}
  V_{\rm Pauli} = 
  \frac{1}{2} 
  C_{\rm P}\left( \frac{1}{q_0 p_0}\right)^3
  \sum_{i, j(\neq i)} 
  \exp{ \left [ -\frac{({\bf R}_i-{\bf R}_j)^2}{2q_0^2} 
          -\frac{({\bf P}_i-{\bf P}_j)^2}{2p_0^2} \right ] }
  \delta_{\tau_i,\tau_j} \delta_{\sigma_i,\sigma_j}\;.
  \label{Pauli}
\end{equation}

In the local potential part, we adopt the Skyrme type with the Coulomb and the symmetry terms as explained in Eq.~(5) of Ref.~\citen{Nii95}, 
\begin{eqnarray}
V_{\rm local} & = & \; {\alpha\over 2\rho_0}\sum_i\langle\rho_i\rangle
        \; + \; {\beta\over (1+\gamma)\;\rho_0^{\gamma}}
                \sum_i\langle\tilde\rho_i\rangle^{\gamma} \nonumber \\
  &   & \; + \; {e^2 \over 2}\sum_{i , j(\neq i)} c_{i} \, c_{j}
          \int\!\!\!\!\int d^3r_i\,d^3r_j 
                 { 1 \over|{\bf r}_i-{\bf r}_j|} \,
                \rho_i({\bf r}_i)\rho_j({\bf r}_j)
                \nonumber \\
  &   & \; + \; {C_{\rm s}\over 2\rho_0} \sum_{i , j(\neq i)} \,
                ( 1 - 2 | c_i - c_j | ) \; \rho_{ij}.
\label{ham1}
\end{eqnarray}
In the above equation, $\rho_0$ is the normal nuclear density ($\simeq 0.165 \rm fm^{-3}$),
$c_i$ is 1 for protons and 0 for neutrons, while $\langle\rho_i\rangle$ and $\langle\tilde\rho_i\rangle$ are overlaps of density with other nucleons defined as
\begin{eqnarray}
\langle\rho_i\rangle     & \equiv & \sum_{j(\neq i)} \; \rho_{ij} \;
              \equiv  \sum_{j(\neq i)}
                     { \int { d^3r \; \rho_i({\bf r}) \;
                       \rho_j({\bf r}) }} \nonumber \\
             & = & \sum_{j(\neq i)}{ (4\pi \lambda^2)^{-3/2}
                  \exp \left[ - ( {\bf R}_i - {\bf R}_j ) ^2
                  / 4\lambda^2 \right] }\;,
\label{rhoij}\\
\langle\tilde\rho_i\rangle     & \equiv & \sum_{j(\neq i)}{ (4\pi \tilde\lambda^2)^{-3/2}
                  \exp \left[ - ( {\bf R}_i - {\bf R}_j ) ^2
                  / 4\tilde\lambda^2 \right] }\;,
\label{rhoijtilde}\\
\tilde\lambda^2 &\equiv& \frac{(1+\gamma)^{1/\gamma}}{2}\lambda^2\;.
\end{eqnarray}

It is known that the nucleon-nucleon interaction has a strong momentum-dependence [see Fig.\ \ref{figPdep}].
We have chosen the form of the momentum-dependent term as a Fock term of the Yukawa-type interaction.
We divide this interaction into two ranges so as to fit the effective mass and the energy dependence of the real part of the optical potential, as
\begin{eqnarray}
V_{\rm MD}  & = & V_{\rm MD}^{(1)} + V_{\rm MD}^{(2)} \nonumber \\
 & = &
         {C_{\rm ex}^{(1)} \over 2\rho_0} \sum_{i , j(\neq i)} 
      {1 \over 1+\left[{{\bf P}_i-{\bf P}_j \over \mu_1}\right]^2} 
      \;\rho_{ij}
  +   {C_{\rm ex}^{(2)} \over 2\rho_0} \sum_{i , j(\neq i)} 
      {1 \over 1+\left[{{\bf P}_i-{\bf P}_j \over \mu_2}\right]^2} 
      \;\rho_{ij}\ .
\label{eqMD}
\end{eqnarray}

Parametrization of the constants in the above effective interactions will be discussed in Sec.~\ref{secParameters}.

%------------------------------------------------------------------------
\subsection{Energy minimum state}
%------------------------------------------------------------------------
For the Hamiltonian with the Pauli potential,
%With inclusion of the Pauli potential, 
we can define the ground state as an energy-minimum state of the system. 
To get the energy-minimum configuration, we use the following damping equations of motion,
\begin{equation}
\dot{{\bf R}}_i =   \frac{\partial H}{\partial {\bf P}_i}
       -\mu_{\bf R} \frac{\partial H}{\partial {\bf R}_i},
\;\;\;\;
\dot{{\bf P}}_i = - \frac{\partial H}{\partial {\bf R}_i}
       -\mu_{\bf P}\frac{\partial H}{\partial {\bf P}_i},
\label{damp00}
\end{equation}
where $\mu_{\bf R}$ and $\mu_{\bf P}$ are the damping coefficients with positive values. %when we need to cool the system.

We first distribute the particles randomly in the phase space and cool down the system according to the damping equations of motion until the energy reaches the minimum value.
Sometimes the system is trapped in a local minimum. % during the cooling.
We thus try again and again this cooling procedure with a different initial state and seek the global energy minimum state.

For finite nuclei and nuclear matter above the saturation density, this procedure works well.
For nuclear matter at subsaturation densities, however, there are many local minimum states around the true ground state, which differ from the ground state in the details of the surface configuration of clusters.
Since the energy difference from the ground state is the order of 10 keV/nucleon in this case, we accept these states as ground states and neglect the small differences of the configuration.

%------------------------------------------------------------------------
\subsection{Periodic boundary conditions}
%------------------------------------------------------------------------

In order to simulate infinite nuclear matter with finite numbers of particles, we use a cubic cell with periodic boundary conditions.
The size of the cell is determined from the average density and the particle number.
The periodic boundary conditions can be introduced as follows:
%% As shown in Fig.~\ref{figCell}, 
  We prepare 26 ($=3^3-1$) surrounding cells, 
which are copies of the central cell.
%%where the particle distribution reflects the distribution of the central cell exactly.
The particles in the central cell move according to the interaction with all particles in the same cell and in the surrounding cells as well.
The particles in the surrounding cells obey exactly the same motions as those in the central cell.
Thus the Hamiltonian per cell is written as
\begin{equation}
H=\sum_{i=1,\cdots, N}
\bigg[\;\; T_i
%  +\sum_{{\rm cell}=0,\cdots, 26 \ } \sum_{j=1,\cdots, N}
  +\sum_{{{\rm cell}=0,\cdots, 26 \ }
          \atop {j=1,\cdots, N}}
     H^{(2)}_{ij}({\bf R}_i-{\bf R}_j+{\bf L}_{\rm cell},\ {\bf P}_i, {\bf P}_j)
  +\cdots\ \ 
\bigg]\ ,
\label{Hamiltonian}
\end{equation}
where $T_i$ is one-body part (kinetic energy), $H^{(2)}_{ij}$ is the two body part of the Hamiltonian and ${\bf L}_{\rm cell}$ are the relative position of surrounding cells from the center.
Note that the indices ``cell'' runs from 0 (the central cell) to 26 (surrounding cells) and ${\bf L}_0=0$.

%------------------------------------------------------------------------
\subsection{Parametrization of the constants}\label{secParameters}
%------------------------------------------------------------------------

We have twelve parameters in the effective interactions of the Hamiltonian (\ref{ham0}), i.e., $C_{\rm P}, q_0, p_0$, $\alpha, \beta, \tau, C_{\rm s}$, $C^{(1)}_{\rm ex}, C^{(2)}_{\rm ex}, \mu_1, \mu_2$, and the Gaussian width $\lambda$. 
We fix these constants to reproduce properties of the ground state of finite nuclei and saturation properties of nuclear matter.

\begin{figure}
\begin{minipage}{0.58\textwidth}
\begin{center}
\includegraphics[width=0.95\textwidth]{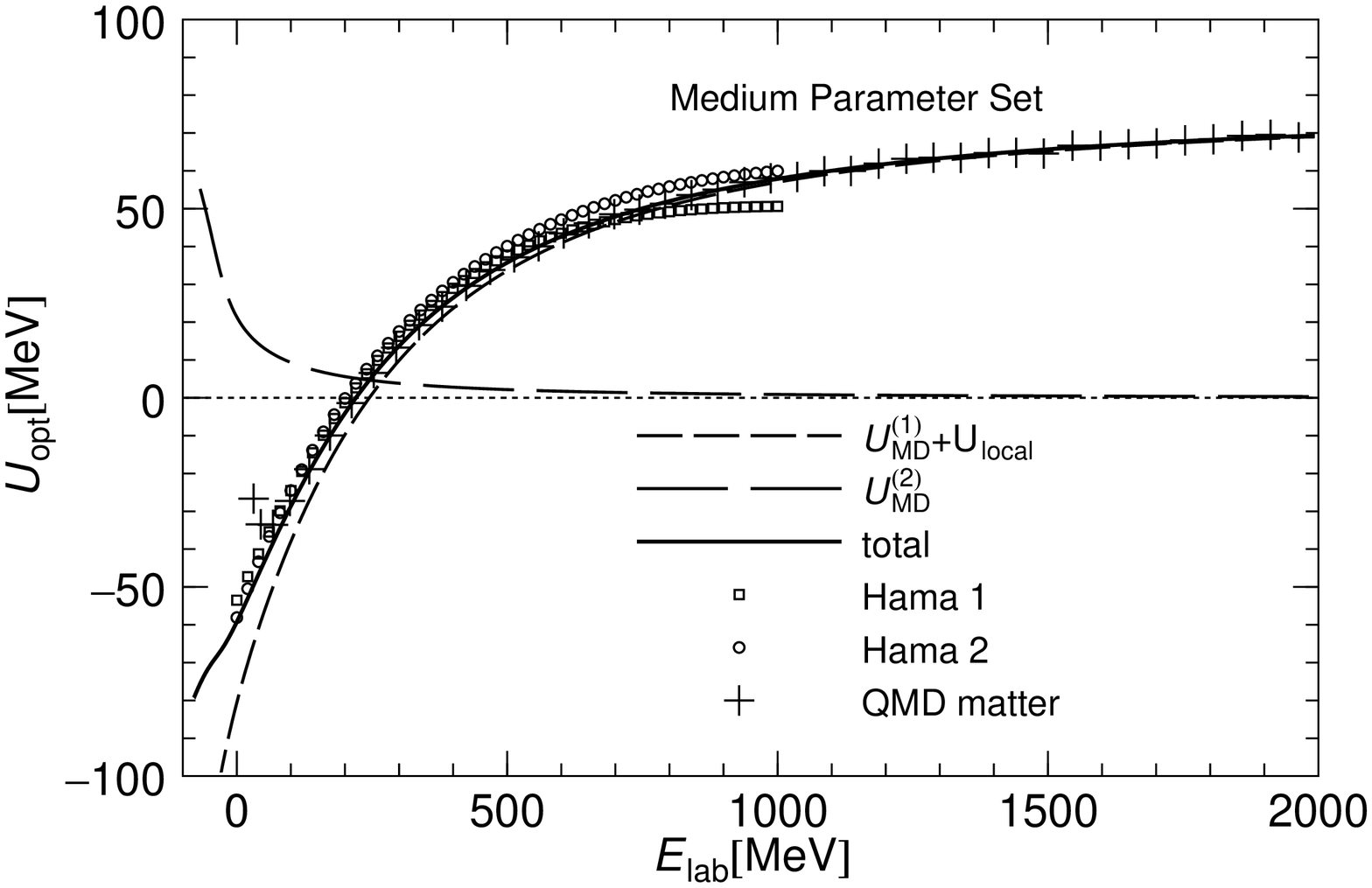}
\caption{Momentum dependence of the potential energy of experimental data and the present QMD model. 
This figure is taken from Ref.\ \citen{QMD-maru}.
\label{figPdep}
}
\end{center}
\end{minipage}
\begin{minipage}{0.4\textwidth}
\begin{center}
\includegraphics[width=0.95\textwidth]{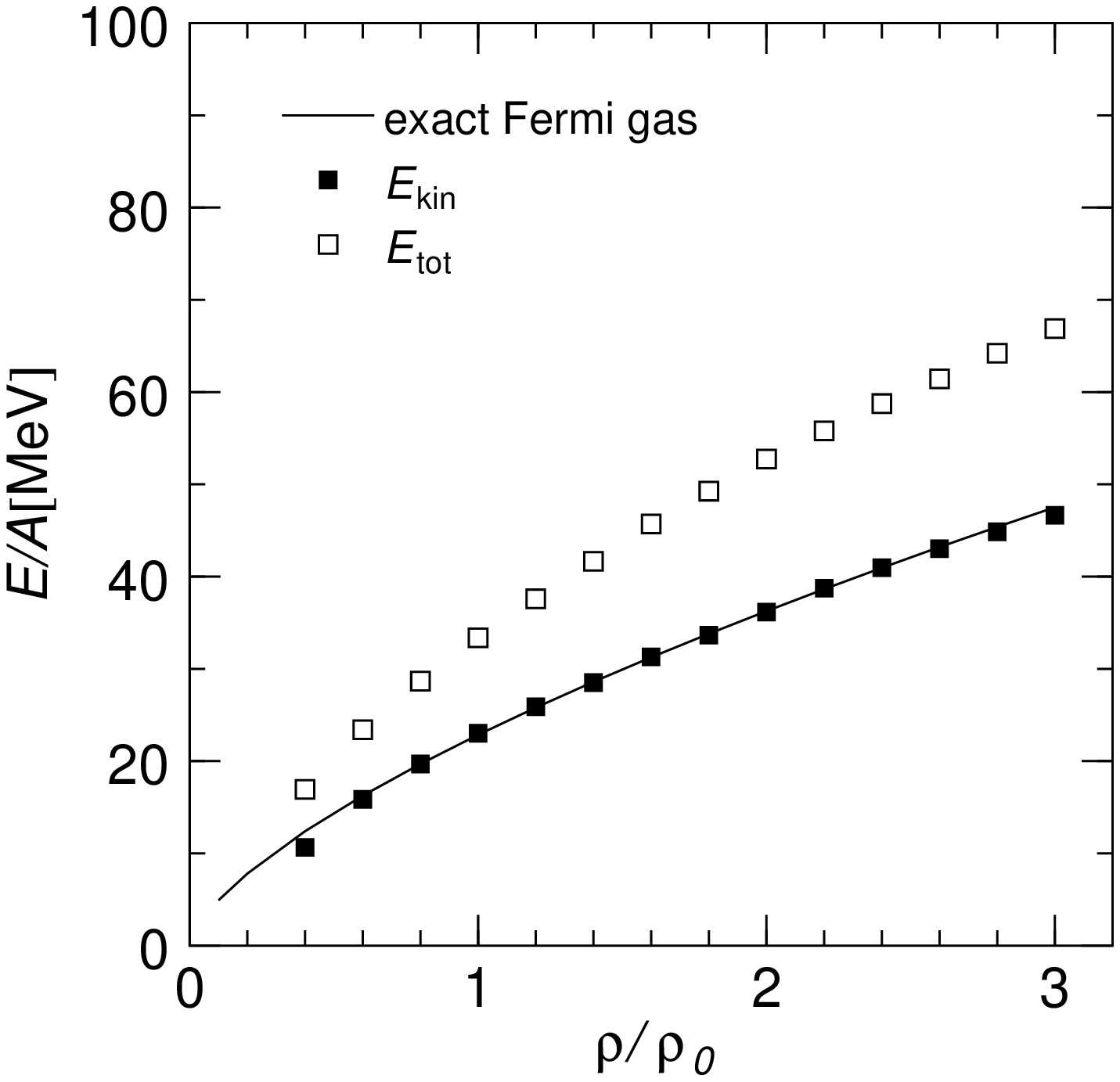}
\caption{
%Density dependence of exact Fermi gas (solid curve) and a system of nucleons which nuclear potentials are switched off and only the Pauli potential is active (filled dots for kinetic energy and open squares for the total energy).
Density dependence of the energy per particle of an ideal Fermi gas.  
The solid curve shows the exact value and the symbols show the QMD results using the Hamiltonian which contains only the kinetic energy and the Pauli potential. 
The filled squares show the kinetic energy and the open squares show the total energy.
This figure is taken from Ref.\ \citen{QMD-maru}.
\label{figPauli}
}
\end{center}
\end{minipage}
\end{figure}

We first determine the parameters, $q_0, p_0$, and $C_{\rm P}$, of Pauli potential apart from the other effective interactions, by fitting the kinetic energy of simulated matter to the energy of the ideal Fermi gas at zero temperature and at various densities.
For this, we define the free Fermi gas system as a ground state for the Hamiltonian including only the kinetic energy and the Pauli potential by making use of the damping equations of motion (\ref{damp00}) and the periodic boundary conditions with 1024 particles in a cell.
In Fig.~\ref{figPauli}, we show the kinetic energy (the solid squares) and the total energy (the open squares) obtained by the Pauli potential with
\begin{equation}
  C_{\rm P}=207\;{\rm MeV},\;\;\; 
  p_0=120\;{\rm MeV},\;\hbox{and}\;\;
  q_0=1.644\;{\rm fm}.
\end{equation}
In the same figure, we plot the energy of the Fermi gas with a solid line.
Although there are some other parameter sets which can reproduce the ideal energies of the Fermi gas using the same form of the Pauli potential, i.e., that used in Ref.~\citen{Pei91}, we choose the above parameter set to get good properties of the ground state of finite nuclei with other effective interaction terms particularly in combination with the momentum-dependent interaction.

Among remaining nine conditions, four are for the momentum-dependent interaction as follows.
We calculate the single particle potential of momentum ${\bf p}$ in nuclear matter at the normal nuclear density, which leads to
\begin{eqnarray}
U({\bf p}, \rho_0) & = & U_{\rm local} + U_{\!\rm MD}({\bf p})
\nonumber  \\
 & = & \alpha  + \beta 
  + \left( \frac{4}{3}\pi p_{\rm F}^3 \right)^{-1}
    \int^{p_{\rm F}} d^3 p'
    \left[{C_{\rm ex}^{(1)} \over 1
  + \left[ {{\bf p}-{\bf p'} \over \mu_1} \right]^2} 
  + {C_{\rm ex}^{(2)} \over 1+\left[ {{\bf p}-{\bf p'} \over \mu_2}
\right]^2} \right] 
\nonumber  \\
 & = & \alpha  + \beta 
    + C_{\rm ex}^{(1)} g(x=\mu_1/p_{\rm F}, y=p/p_{\rm F})
    + C_{\rm ex}^{(2)} g(x=\mu_2/p_{\rm F}, y=p/p_{\rm F}),
\nonumber  \\
\label{eqUOpt}
\end{eqnarray}
with
\begin{equation}
g(x,y) = \frac{3}{4} x^3 \left[ 
         \frac{1+x^2-y^2}{2xy}
         \ln{ \frac{(y+1)^2+x^2}{(y-1)^2+x^2}}
       + \frac{2}{x}
       - 2 \left\{ \arctan{ \frac{y+1}{x} }
                - \arctan{ \frac{y-1}{x} }
           \right\} \right] .
\end{equation}
We fit the energy dependence of this potential to experimental data.
In Fig.~\ref{figPdep}, we plot the energy dependence of the real part of the optical potential (the open circles and squares) obtained from the experimental data of Hama \etal \cite{Ham90} for proton-nucleus elastic scattering.
% From this figure, we pick up three conditions, 
To fit the data, we impose three constraints, i.e., $U(0) = -80$ MeV, $U(p)=0$ at $E_{\rm lab} = 200$ MeV, and $U(p \rightarrow \infty) = \alpha + \beta = 77$ MeV.
%For another condition, we use the value of effective mass defined by
For another condition, 
we take the effective mass $m^*=0.8\,m$ at $\rho=\rho_0$, where
\begin{equation}
 \frac{1}{m^*} = \frac{1}{m} +
 \left( \frac{1}{p} 
 \frac{\partial U_{\!\rm MD}}{\partial p} 
 \right)_{p=p_{\rm F}}.
\end{equation}
%We take $m^*=0.8\,m$ at $\rho=\rho_0$.

Other three conditions are coming from the saturation condition, i.e., the energy per nucleon $E/A=-16$ MeV and the incompressibility $K$ at $\rho=\rho_0$.
%In the present model, we set the saturation density $\rho_0=0.165$ fm$^{-3}$.

\begin{figure}
\begin{center}
\includegraphics[width=0.65\textwidth]{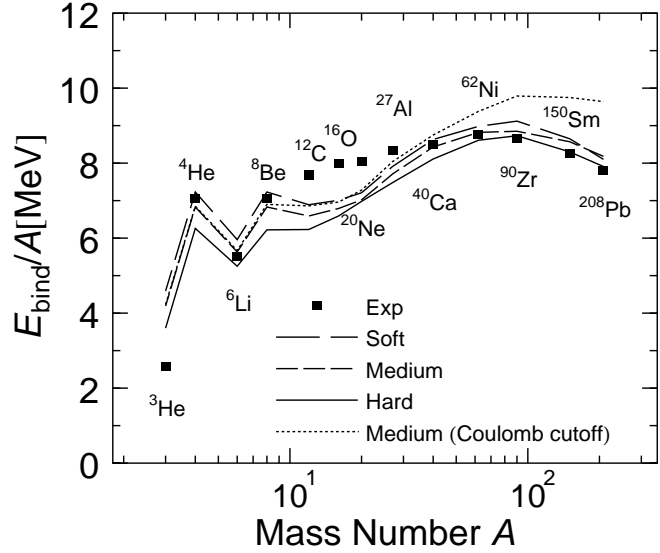}
\caption{Binding energies of nuclei calculated with our QMD model.
	There are three sets of parameters, soft, medium, and hard, which refer to the stiffness of nuclear matter at the saturation.
This figure is taken from Ref.\ \citen{QMD-maru}.
\label{figEbindNucl}
}
\end{center}
\end{figure}

There are two parameters left.
One is the symmetry energy coefficient $C_{\rm s}$, which we take %to be 
$25$ MeV to get a reasonable value of the symmetry energy $34.6$ MeV for nuclear matter at the saturation density.
The other is the width of the Gaussian wavepacket $\lambda$, 
%which is a free parameter in QMD model, 
%This value affects 
which is chosen to get appropriate ground state properties of finite nuclei and infinite nuclear matter below the saturation density 
%while it does not change those of uniform nuclear matter above the saturation density.
without changing those of uniform nuclear matter above the saturation density.
We then choose these parameters to give a good fitting to the binding energies of finite nuclei plotted in Fig.\ \ref{figEbindNucl}.

\begin{figure}
\begin{center}
\includegraphics[width=0.85\textwidth]{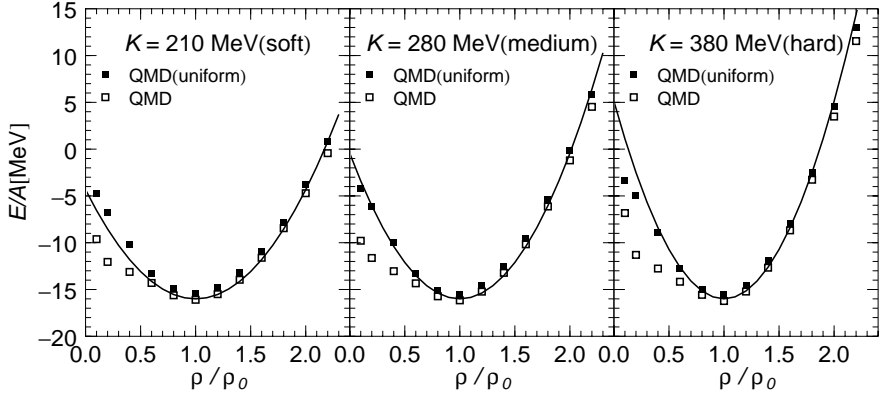}
\caption{Energy per nucleon calculated with our QMD model.
This figure is taken from Ref.\ \citen{QMD-maru}.
\label{figRhoE}
}
\end{center}
\end{figure}

It should be noted here that we cannot determine these parameters from the above conditions in an analytical way, since the Fermi distribution is not exactly achieved by the Pauli potential and the additional potential energy included in the Pauli potential.
%%In addition, the saturation properties of nuclear matter should be realized in the simulated matter for the main purpose of this paper.
Thus we  simulate nuclear matter by QMD with the periodic boundary conditions using 1024 particles in a cell. 
We search the energy minimum state by the damping equations of motion (\ref{damp00}) as discussed above and adjust the parameters.
%%By this method, we have fixed three parameter sets with different values of incompressibility $K$, which are shown in Table \ref{table1c}. 
%%We have prepared three equations of state (EOSs), namely Soft ($K$=210 MeV), Medium ($K$=280 MeV), and Hard ($K$=380 MeV) EOSs.
By this method, we have fixed three parameter sets 
corresponding to equations of state (EOSs)
with different values of incompressibility $K$:
Soft ($K$=210 MeV), Medium ($K$=280 MeV), and Hard ($K$=380 MeV) EOSs \cite{QMD-maru}. 
These values of the incompressibility $K$ are extracted from the curvature of the energy per nucleon (shown in Fig. \ref{figRhoE}) at the saturation density by fitting to the following parabolic form,
\begin{equation}
 E/A=\frac{K}{18\rho_0^2}\left(\rho-\rho_0\right)^2-16\;\;{\rm MeV}\;.
\end{equation}
%
%This parabola is also shown in Fig.~\ref{figRhoE}.

The single particle potential shown in Fig.~\ref{figPdep} are also calculated by the simulated nuclear matter with the Pauli potential with the effective interactions (\ref{ham0}). 
%The results are denoted by the crosses in Fig.~\ref{figPdep} and agree well with the results of ideal nuclear matter except for the low energy part, where the Pauli potential is effective.
The results are denoted by the crosses in Fig.~\ref{figPdep} and agree well with the single particle potential given by Eq.\ (\ref{eqUOpt}) except for the low energy part, where the Pauli potential is effective.
Though this result in Fig.~\ref{figPdep} is obtained with the parameter set of Medium EOS, results with Soft and Hard EOSs are the same as Medium EOS within 2 MeV for the whole energy region.

In Fig.~\ref{figEbindNucl}, we plot the binding energies of the ground state of finite nuclei obtained by the damping equations of motion (\ref{damp00}) with three parameter sets, i.e., Soft (the long dashed line), Medium (the dashed line), and Hard (the solid line) EOSs.
All of them describe well the global trend of the binding energies of various nuclei except for light nuclei from $^{12}$C to $^{20}$Ne.
The disagreement for the light nuclei might be due to the individual structures (shell structures, cluster structures, etc.) of these light nuclei, which are not well described by the present QMD.
The parameters for Medium EOS which we use for our QMD calculations are listed in Table \ref{table1b}.

\begin{table}
\begin{center}
\caption{\label{table1b}
Effective interaction parameter set ($K$=280 MeV).}
%\begin{tabular}{|cc|cc|c|cc|}
\begin{tabular}{ cc cc c cc }
\hline
& $\alpha$ (MeV) & $-92.86$ & 
& $\beta$ (MeV) & 169.28 & \\
& $\gamma$ &$4/3$&
& $C_{\rm s0}$ (MeV) &$25.0$&\\
& $C_{{\rm ex}(1)}$ (MeV) &$-258.5$&
& $C_{{\rm ex}(2)}$ (MeV) &$375.6$&\\
& $\mu_1$ (MeV) &$2.35$&
& $\mu_2$ (MeV) &$0.4$&\\
& $C_{\rm P}$ (MeV) &$115.0$&\\
& $p_0$ (MeV) &$207.0$&
& $q_0$ (fm) &$1.644$&\\
%& $L$ (fm$^2$) &$2.15$&
& $\lambda$ (fm) &$1.45$&
&              &      &\\          
\hline
\end{tabular}
\end{center}
\end{table}

\subsubsection{Electron background}

The total charge of the system should be zero, i.e., the numbers of electrons and protons in the cell are equal.
%As the size of the cell 
As the distance between nuclei
is small compared with the Thomas-Fermi screening length by electrons $\lambda_{\rm TF}^e$, where
\begin{eqnarray}
4\pi{\lambda_{\rm TF}^e}^2&=& {\frac{\partial\mu_i}{\partial\rho_{e}}},
\end{eqnarray}
the distribution of electrons should be almost uniform.
In fact, the inhomogeneity of electron distribution is found to be small by our studies \cite{screening,Maruyama05} which include the screening by electrons.

Therefore, it is natural to treat electrons as a uniform background.
In this case, we should take account of the long-range contributions of the Coulomb interaction between protons.
However, for simplicity a cutoff of the Coulomb interaction, e.g., in the form of screened Coulomb interaction, is sometimes used \cite{Pei91,Horowitz04,QMD-maru}.
In Sec.\ \ref{QMDmatter-maru} we employ this ``screened'' Coulomb interaction among protons, 
\begin{equation}
      V_{\rm C}^{\rm scr} =
    {e^2 \over 2}\sum_{i , j(\neq i)} c_{i} \, c_{j}
     \int\!\!\!\int d^3r_i\,d^3r_j { 
    \exp\left[-|{\bf r}_i-{\bf r}_j|/r_{\rm cut} \right] 
              \over|{\bf r}_i-{\bf r}_j|} \,
           \rho_i({\bf r}_i)\rho_j({\bf r}_j)\;,
\end{equation}
where $r_{\rm cut}$ is the cutoff length,  which we set 10 fm to cutoff the interaction within the length of the cell size.
The physical screening length of the Coulomb potential by the electron localization is, however, estimated to be much larger in the case of the normal nuclear density \cite{Williams85}.
Thus our ``screening'' should be considered as a technical approximation to avoid this cell-size dependence and to make the numerical calculation feasible.
In Sec.\ \ref{QMDmatter-wata} on the other hand, we employ completely uniform electron background and fully include the long-range Coulomb interaction of protons in replica cells by the Ewald summation method.
In both the cases we do not simulate electrons explicitly by QMD.

\section{Nuclear matter at subsaturation densities by QMD\label{sec-matter}}
\subsection{Appearance of inhomogeneous structures in nuclear matter\label{QMDmatter-maru}}

Through a number of works, it has been clarified that the pasta phases might be the ground state of matter at subsaturation densities for various nuclear interactions including typical ones \cite{Ravenhall83,Hashimoto84,Lassaut87,Williams85,Lorenz93,Maruyama05,Oyamatsu93,Sumiyoshi,Watanabe00,Magierski,Oyamatsu2007,Gogelein2007,Avancini,Newton09}.  
%XXXXXX HERE WE CAN CITE MANY PAPERS IF WE WANT XXXXXX
In addition, the pasta phases can occupy a significant mass fraction of neutron star crusts ($\simeq 50$\percent)\cite{Lorenz93} and collapsing supernova cores ($\gtrsim 20$\percent)\cite{opacity} if they really exist in these objects.

However, almost all the previous works assume several possible shapes of nuclei and what they can actually claim is that the pasta phases can be the energetically most favorable state among the selected specific structures.
Furthermore, all previous studies are based on static frameworks and focus only on the equilibrium state, mainly the ground state.  
Therefore, the fundamental problem whether or not the pasta phases are actually formed in young neutron stars in their cooling process and supernova cores in the stage of the gravitational collapse has been totally unclear.

To solve this problem, we have studied whether the pasta phases are formed by adiabatically changing an external parameter (either decreasing the temperature or increasing the density) without any assumption on the nuclear shape.  \cite{QMD-wata-rapid,QMD-wata,QMD-finiteT-wata,QMD-sono,QMD-wata-transition,QMD-wata-formation}
For this purpose, QMD which enables us to simulate the time evolution of the nucleon many-body systems with a large number of nucleons is very powerful.
It is also noted that we are mainly interested in the nuclear structure from the mesoscopic to macroscopic scales of $\gtrsim 10$ fm, where the exchange effect should be less important.
Therefore, it is expected that QMD is a reasonable approximation for studying the pasta phases.
%Especially, at non-zero temperatures of $\gtrsim O(1)$ MeV, validity of QMD is ensured because the shell structure caused from the exchange effect is washed out by thermal fluctuations above $T\sim 3$ MeV \cite{Newton09}.
Especially, at non-zero temperatures of $\gtrsim O(1)$ MeV, validity of QMD is ensured because the shell effects are washed out by thermal fluctuations above $T\sim 3$ MeV \cite{Newton09}.

We consider a system with neutrons, protons, and electrons in a cubic box with periodic boundary conditions.  
The system is not magnetically polarized, i.e., it contains equal numbers of protons (and neutrons) with spin up and spin down.
Relativistic degenerate electrons which ensure charge neutrality can be regarded as a uniform background because electron screening is negligible at relevant densities around the normal nuclear density \cite{Pethick95,screening,Maruyama05} as we have discussed in the last section.
If we assume completely uniform electron distribution, we have to take account of the long-range nature of the Coulomb interaction. 
First, we show the results of our QMD calculations with a cutoff distance of the Coulomb interaction, in order that the range of the interaction does not exceed the cell size \cite{QMD-maru}.
Similar calculation has been done also in a pioneering work by Peilert \etal using a different QMD model \cite{Pei91}.
\footnote{
Recently, Horowitz and his collaborators have also studied the structure of nuclear matter at subsaturation densities using QMD \cite{Horowitz04,Horowitz04b}.
Their model is close to the early version of QMD without the Pauli potential, and thus they cannot simulate systems at zero temperature.}

\begin{figure}[tbp]
\begin{center}
{\includegraphics{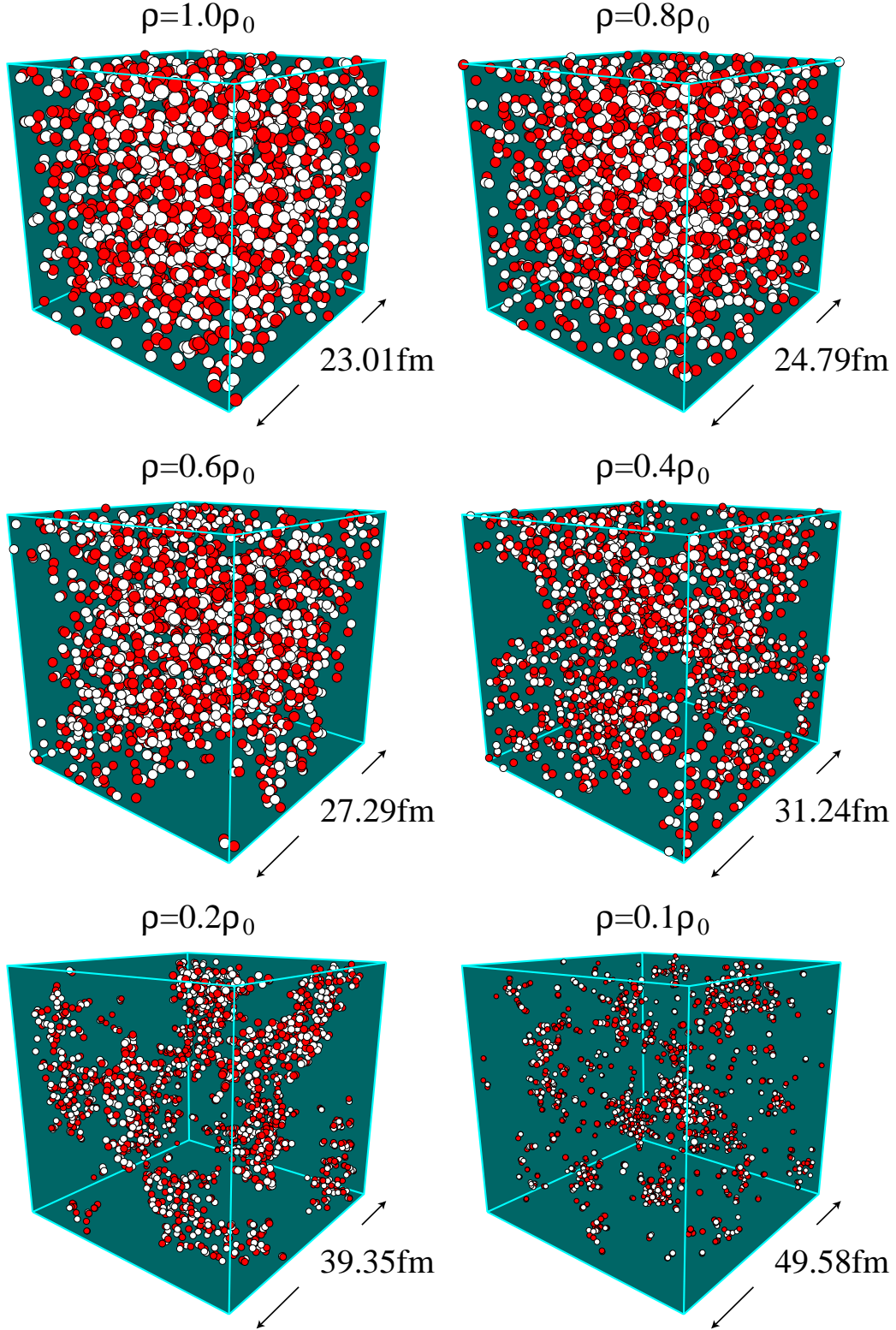}}
\caption{\label{figsnap1-maru}
%% Figure \ref{figsnap1-maru} shows snapshots of symmetric nuclear matter at various densities.
%% Below $0.6\rho_0$, the ground state of matter becomes inhomogeneous \cite{Pei91}.
%% We can see structures similar to nuclear ``pasta''.
%% However, these structures are not regular and some of them are hard to classify into typical pasta structures.
%% We consider this is due to the use of a cutoff distance for the Coulomb interaction and possibly due to too rapid cooling.
%% 
  Nucleon distributions of symmetric nuclear matter $x=0.5$ at $T\simeq 0$.
  The total number of nucleons in the simulation is $1024$.
  The red particles show protons and the white ones show neutrons.
  For densities above $0.8\rho_0$, matter is uniform.
  At lower densities, there appear some incomplete pasta-like structures:
  spherical bubbles ($0.6\rho_0$), rod-like nuclei ($0.2\rho_0$), and spherical nuclei ($0.1\rho_0$). 
This figure is taken from Ref.\ \citen{QMD-maru}.
  }
\end{center}
\end{figure}

\begin{figure}[tbp]
\begin{center}
{\includegraphics{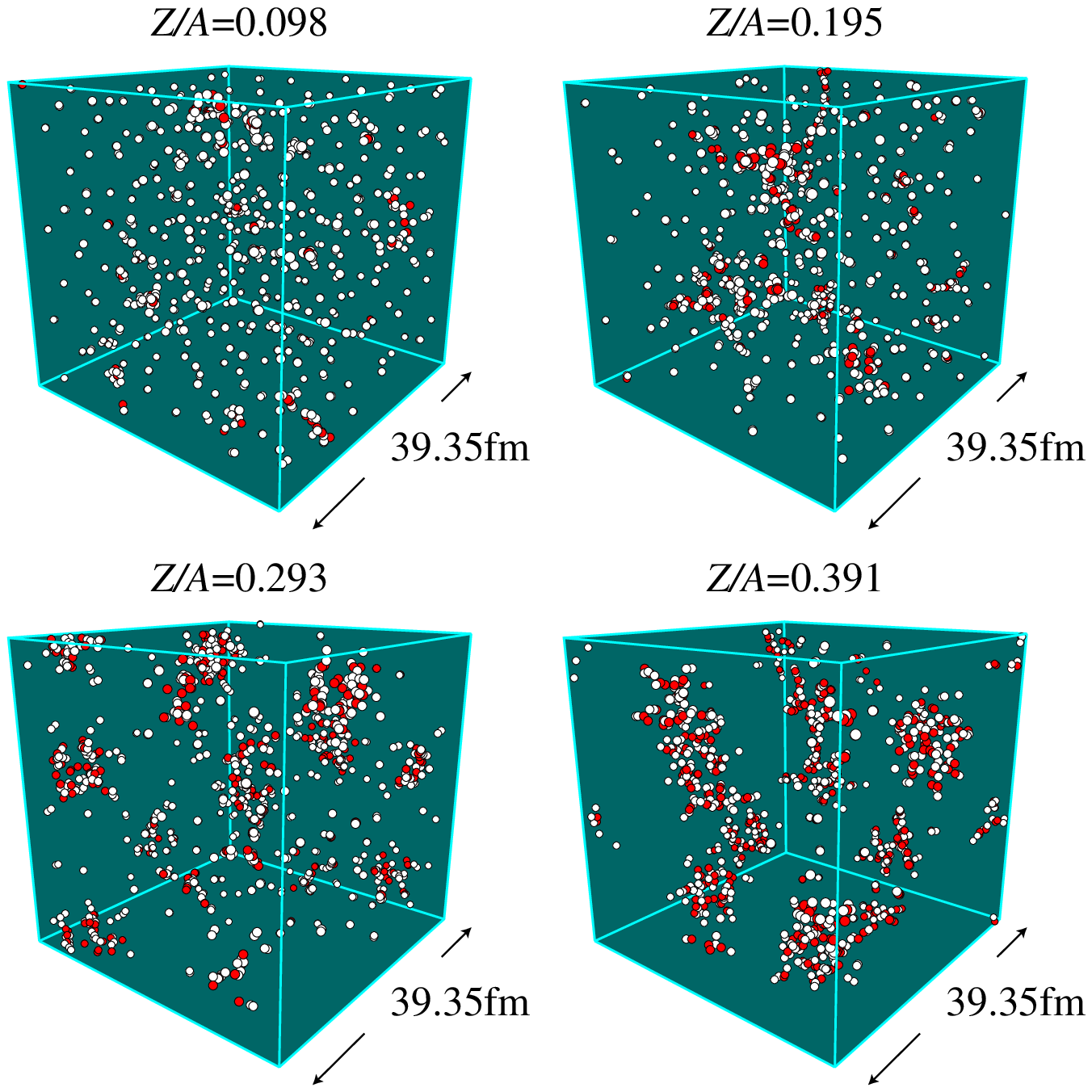}}
\caption{\label{figsnap2-maru}
  Nucleon distributions of asymmetric nuclear matter for $\rho=0.1\rho_0$ at $T\simeq 0$.
  The dependence on the proton fraction $x=Z/A$ is shown.
This figure is taken from Ref.\ \citen{QMD-maru}.
  }
\end{center}
\end{figure}

Figure \ref{figsnap1-maru} is snapshots of symmetric nuclear matter at various densities.
Below $0.6\rho_0$, the ground state of matter becomes inhomogeneous \cite{Pei91}.
We can see structures similar to nuclear ``pasta''.
However, these structures are not regular and some of them are hard to classify into typical pasta structures.
In addition, we could not realize the bcc lattice of spherical nuclei at low densities of $\sim 0.1\rho_0$, which must be the true ground state in this density region.
\footnote{
Very recently, one of the present authors and his collaborators have found that fcc lattice of spherical nuclei can be the ground state, by taking the optimum sizes of the cell and nuclei as well as the inhomogeneous electron distribution \cite{Okamoto}.}
We consider the reasons of these incompleteness and why we could not obtain the pasta phases are the use of a cutoff distance for the Coulomb interaction and possibly due to too rapid cooling.

Energy per nucleon of symmetric nuclear matter is plotted by open squares in Fig.\ \ref{figRhoE}.
The contribution of electrons is subtracted and only the nucleon contribution is included.
Compared with uniform distribution (filled squares), the energy gets lower at densities below $\sim0.6\rho_0$.
This is due to the formation of inhomogeneous structures.

Figure \ref{figsnap2-maru} shows the dependence of the structure on the proton fraction.
If the proton fraction $x$ is close to 0.5, neutrons and protons distribute congruently \cite{NS_review}.
However, with decrease of $x$, extra neutrons drip out of the nuclei and spread the whole space.

%\subsection{Nuclear matter at subsaturation densities with full Coulomb interactions\label{QMDmatter-wata}}
\subsection{Pasta formation by decreasing temperature\label{QMDmatter-wata}}

%Though we have shown that inhomogeneous structures appear and has significant effects on the EOS, it is still unclear whether pasta structures actually appear or not.
%
The above QMD results confirm that inhomogeneous structures appear at subsaturation densities and this phenomenon has a significant effect on the EOS.  
However, an important question whether or not the pasta phases actually appear is still unclear.  
To solve this problem, we have performed another series of QMD studies\cite{QMD-wata-rapid,QMD-wata,QMD-finiteT-wata,QMD-sono} in which we include the long-range contribution of the Coulomb interaction and perform careful simulated annealing to achieve thermal equilibrium.  
Here we calculate the Coulomb interaction by the Ewald summation method, which enables us to sum up the contributions of long-range interactions in a system with periodic boundary conditions efficiently.
%For nuclear interaction, we used the QMD Hamiltonian of Ref.\ \citen{QMD-maru} with the standard medium EOS parameter set and that of Refs.\ \citen{Kido,Chikazumi00,Chikazumi01}.
For nuclear interaction, we use the QMD Hamiltonian of Ref.\ \citen{QMD-maru} with the standard medium EOS parameter set and another form of the QMD Hamiltonian of Refs.\ \citen{Kido,Chikazumi00,Chikazumi01}.
The qualitative results are the same for both the models \cite{QMD-sono}.

\begin{figure}[tbp]
\begin{center}
%\resizebox{14cm}{!}
%{\includegraphics{uni_x0.3_label_2.ps}}
{\includegraphics[width=0.85\textwidth]{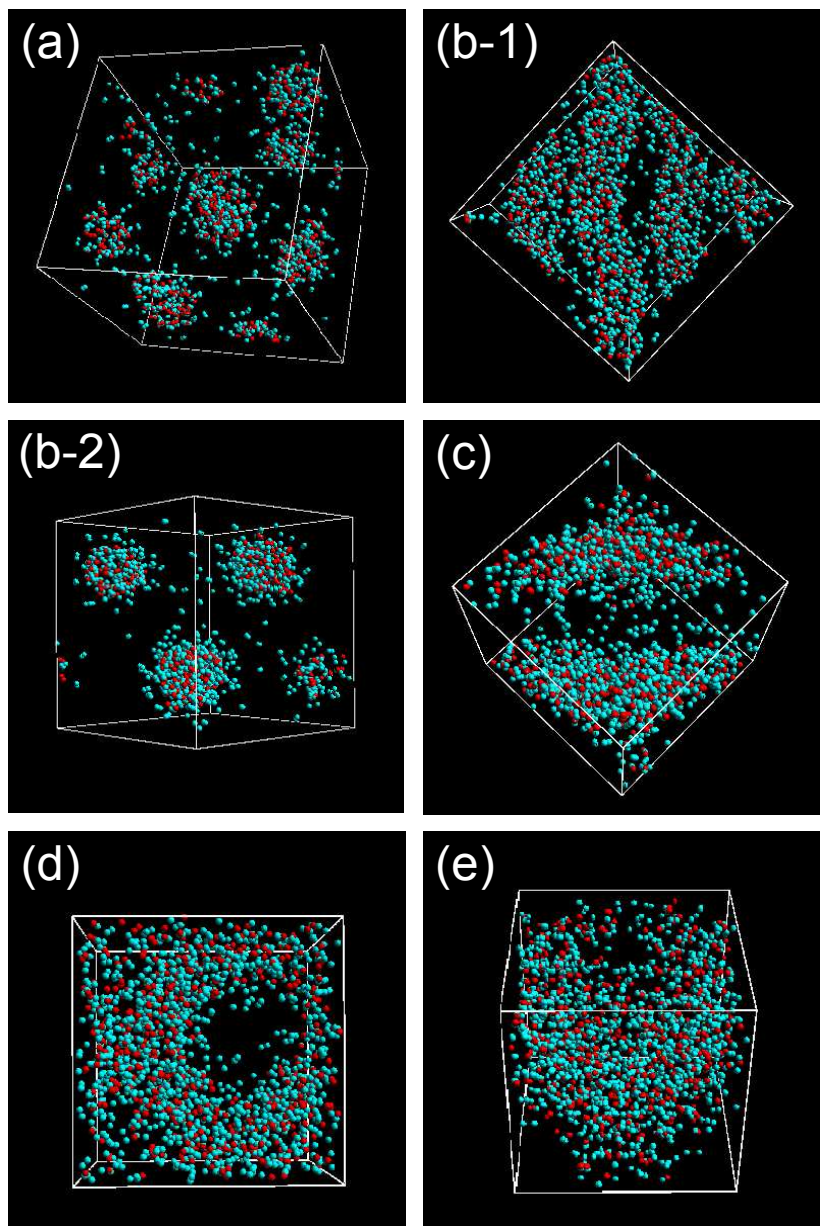}}
\caption{\label{fig_x0.3_t0}
  Nucleon distributions of the pasta phases for $x=0.3$ at $T\simeq 0$ MeV.
  The total number of nucleons in the simulation is $2048$ ($614$ protons
  and $1434$ neutrons).
  The red particles show protons and the green ones show neutrons.
  Each panel shows the pasta phase with (a) spherical nuclei ($0.1\rho_0$),
  (b-1) rod-like nuclei ($0.18\rho_0$); side view, (b-2) the same; top view, 
  (c) slab-like nuclei ($0.35\rho_0$),
  (d) rod-like bubbles ($0.5\rho_0$), and (e) spherical bubbles ($0.55\rho_0$).
  This figure is adapted from Ref.\ \citen{QMD-wata}.
  }
\end{center}
\end{figure}

In Fig.\ \ref{fig_x0.3_t0}, we show the resulting snapshots of the nucleon distributions for $x=0.3$ at $T\simeq 0$ MeV.
Note that we obtain the pasta structures without assuming the nuclear shapes {\it a priori} \cite{QMD-wata-rapid,QMD-wata}.
This is the first result which shows the formation of the pasta phases using dynamical framework.
In the simulations, we first prepare a uniform hot nucleon gas at $T\sim 20$ MeV for each density. Starting from this initial condition, we slowly cool it down using frictional relaxation method, which is given by QMD equations of motion plus small friction terms [Eq.\ (\ref{damp00})].
%%\begin{equation}
%%  \dot{\bf R}_{i} =
%%     {\displaystyle \frac{\partial {\cal H}}{\partial {\bf P}_{i}}
%%       - \xi_{R} \frac{\partial {\cal H}}{\partial {\bf R}_{i}}}\ ,\qquad
%%  \dot{\bf P}_{i} =
%%     {\displaystyle -\frac{\partial {\cal H}}{\partial {\bf R}_{i}}
%%       - \xi_{P} \frac{\partial {\cal H}}{\partial {\bf P}_{i}}}\ ,\label{qmdeom fric}
%%\end{equation}
%%where $\xi_{R}$ and $\xi_{P}$ are the positive friction coefficients.
Throughout this cooling process, we keep the quasi-thermal equilibrium.
We take the time scale of $O(10^{3}-10^{4})$ fm/$c$ to reduce the temperature down to $\sim 0.1$ MeV or less.
This result suggests that the pasta phases can be formed in the neutron star crusts by cooling.
It should be noted, however, that the typical value of the proton fraction $x$ in the relevant region of neutron star crusts is $\lesssim 0.1$, which is lower than that used in these simulations.
However, we have also done similar simulations at $x=0.1$ and have observed the formation of the pasta phases \cite{QMD-wata}.  

%Non-zero temperatures

\begin{figure}[tbp]
\begin{center}
\resizebox{14cm}{!}
{\includegraphics{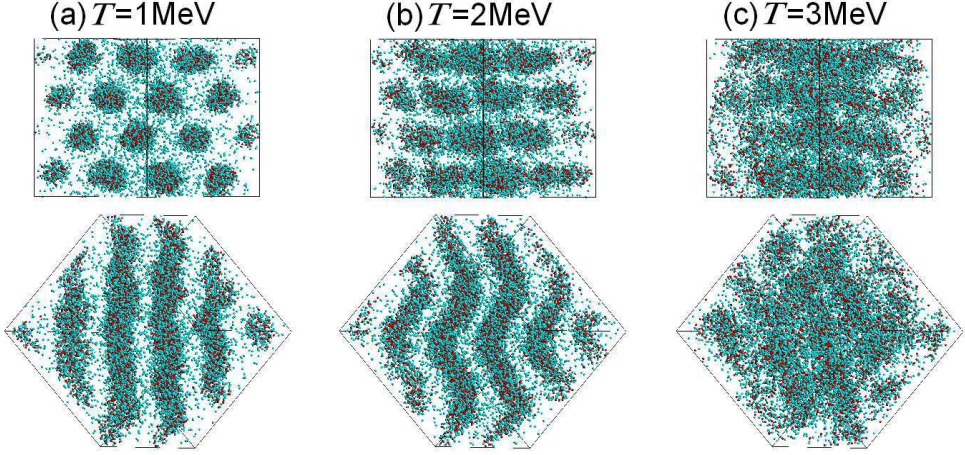}}
\caption{\label{fig_rod}
  Nucleon distributions at $T=1$, 2, and 3 MeV 
  for $x=0.3$ and $\rho=0.175\rho_{0}$, where the phase with rod-like nuclei
  is obtained at zero temperature.
  The total number of nucleons in this simulation is $16384$ 
  ($4915$ protons and $11469$ neutrons).
  The upper figures show top views along the axis of the rod-like nuclei 
  at $T = 0$, and the lower ones show side views.
  The red particles represent protons and green ones represent neutrons.
  This figure is taken from Ref.\ \citen{QMD-finiteT-wata}.
  }
\end{center}
\end{figure}

\begin{figure}[tbp]
\begin{center}
\resizebox{14cm}{!}
{\includegraphics{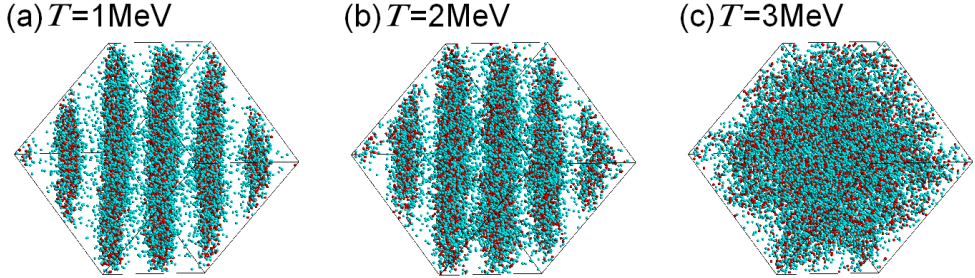}}
\caption{\label{fig_slab}
  The same as Fig.\ \ref{fig_rod} for $x=0.3$ and $\rho=0.34\rho_{0}$,
  where the phase with slab-like nuclei is obtained at zero temperature.
  The line of sight of these figures is in the direction parallel
  to the plane of the slab-like nuclei at $T=0$.
  This figure is taken from Ref.\ \citen{QMD-finiteT-wata}.
  }
\end{center}
\end{figure}

Another important advantage of QMD is that effects of non-zero temperature can be naturally incorporated.
Therefore, QMD simulations provide us a clear picture how the pasta structures shown in Fig.\ \ref{fig_x0.3_t0} are formed by decreasing temperature \cite{QMD-finiteT-wata}.
%In Figs.\ \ref{fig_rod} and \ref{fig_slab}, we show snapshots of the nucleon distributions at non-zero temperatures for $x=0.3$ and $\rho=0.175 \rho_0$ and $\rho=0.34 \rho_0$, respectively; at $T\simeq 0$, we obtain the pasta phase with rod-like nuclei in the former case, and that with slab-like nuclei in the latter case.
In Figs.\ \ref{fig_rod} and \ref{fig_slab}, we show snapshots of the nucleon distributions at non-zero temperatures for 
   $x=0.3$ at $\rho=0.175\rho_0$ and $0.34\rho_0$,
  %$x=0.3, \rho=0.175 \rho_0$ and $x=0.3, \rho=0.34 \rho_0$, 
  respectively; at $T\simeq 0$ MeV, we obtain the pasta phase with rod-like nuclei in the former case, and that with slab-like nuclei in the latter case.
Around these densities, the phase separation occurs at $T\sim 5$ MeV and we see that, at $T\sim 3$ MeV, the density inhomogeneity by clustering of nucleons becomes significant [Figs.\ \ref{fig_rod}(c) and \ref{fig_slab}(c)].
At $T\simeq 2$ MeV, nuclear shapes become recognizable even though the surface diffuseness of nuclei and the fluctuation of the nuclear shape are still large and there are still many evaporated nucleons among nuclei [Figs.\ \ref{fig_rod}(b) and \ref{fig_slab}(b)].
By further decreasing temperature, these surface diffuseness, fluctuation of the nuclear shape, and the number of evaporated nucleons except for dripped neutrons become small and, eventually, clear pasta structures can be observed at $T\lesssim 1$ MeV [Figs.\ \ref{fig_rod}(a) and \ref{fig_slab}(a)].

\begin{figure}[t]
\begin{center}
\rotatebox{270}{
\resizebox{9.3cm}{!}
{\includegraphics{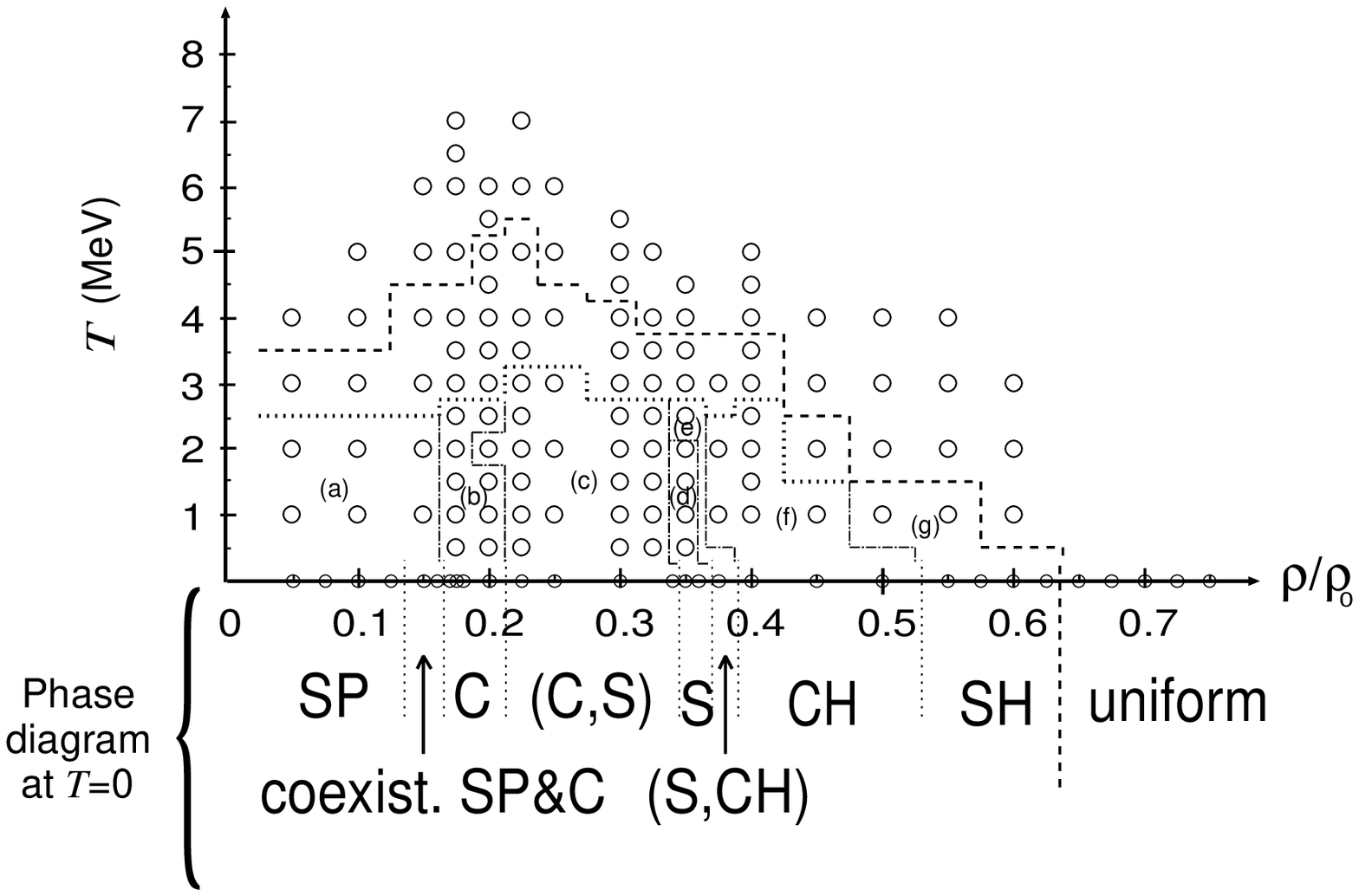}}}
\caption{\label{phase diagram x0.3}
  Phase diagram of matter at $x=0.3$ plotted in the $\rho$ -- $T$
  plane.  The dashed lines correspond to phase separation lines. The
  thick dotted lines show the boundary above which nuclear surface cannot be
  identified. The dashed-dotted lines show boundaries between different
  phases.  Abbreviations SP, C, S, CH, and SH mean phases with
  spherical nuclei, cylindrical (rod-like) nuclei, slab-like nuclei, cylindrical
  holes, and spherical holes, respectively. The parentheses (A,B) show
  an intermediate phase between A and B phases.  Simulations have been
  carried out at the points denoted by circles.  
  This figure is adapted from Ref.\ \citen{QMD-finiteT-wata}.
  }
\end{center}
\end{figure}

Finally, we summarize our results in a phase diagram at subsaturation densities shown in Fig.\ \ref{phase diagram x0.3}.
In the region below the thick dotted lines, where we can identify the nuclear surface, we have obtained the pasta phases with spherical nuclei [region (a)], rod-like (cylindrical) nuclei [region (b)], slab-like nuclei [region (d)], cylindrical holes [region (f)] and spherical holes [region (g)].
It is noted that, in addition to the pasta phases of these simple structures, phases with more complicated structures whose both the nuclear matter region and the bubble region have multiply-connected configurations have been obtained [regions (c) and (e)].
Existence of phases with such complicated structures, e.g., gyroid and double-diamond phases, have been discussed by several authors using different methods \cite{Williams85,Lassaut87,Nakazato09,Okamoto,Pethick95}.


\subsection{Pasta formation by compression of nuclear matter\label{sect_qmdsn}}
%\subsection{Compression of nuclear matter and dynamical change of structure}

In supernova cores, pasta phases are expected to be formed by compression of matter in the gravitational collapse [see, e.g., Refs.\ \citen{soft_review,pasta_review} and references therein].
This issue is, however, more non-trivial compared to the pasta formation by decreasing temperature discussed in Sec.\ \ref{QMDmatter-wata} because drastic changes of the nuclear structure, such as from sphere to rod, must be involved in the present case.
Therefore, to solve whether or not the pasta phases are formed in collapsing supernova cores and to understand its formation process, an {\it ab-initio} dynamical approach is needed.
%;
Using QMD, which is very suitable for this purpose, we have solved this problem by demonstrating that a lattice of rod-like nuclei is formed from a bcc lattice of spherical nuclei by compression  \cite{QMD-wata-formation}.%
\footnote{We have shown that a layered lattice of slab-like nuclei and triangular lattice of cylindrical bubbles are also formed by further compression \cite{QMD-wata-transition}}.
Our results establish that the pasta phases can be formed in collapsing supernova cores.

A generally accepted scenario of the formation of the pasta phases in supernova cores is that, when the density exceeds some critical value, an instability of nuclear fission sets in and, consequently, all the nuclei elongate in the same direction and ``eventually join up to form string-like structures''\cite{Pethick95}.
When the Coulomb energy between protons in a nucleus is sufficiently larger than the surface energy of the nucleus, the reduction of the Coulomb energy due to the fission exceeds the energy cost of the surface tension by an increase of the surface area, and the fission barrier vanishes (i.e., the onset of the fission instability).
This is described by the celebrated Bohr-Wheeler condition:
\begin{equation}
  E_{\rm Coul}^{(0)} \geq 2 E_{\rm surf},
\label{eq_bohr_wheeler}
\end{equation}
where $E_{\rm Coul}^{(0)}$ and $E_{\rm surf}$ are the self-Coulomb and surface energies of the nucleus.
Using an equilibrium condition between the Coulomb and surface energies evaluated within the WS approximation, one can show that Eq.\ (\ref{eq_bohr_wheeler}) reads $u\geq 1/8$, where $u$ is the volume fraction occupied by nuclei \cite{Pethick95}.

However, we should note that Bohr-Wheeler condition (\ref{eq_bohr_wheeler}) is derived for an isolated nucleus in vacuum.
In the real situation in supernova cores, there are background electrons and the condition for the fission instability should be modified from the original Bohr-Wheeler condition (\ref{eq_bohr_wheeler}).
Indeed, it has been shown that the fission instability is suppressed by the background electrons which reduce the local net charge density inside nuclei \cite{Brandt,Burvenich07}.
This result poses a doubt about the formation scenario based on the fission instability.

\begin{figure}[tbp]
\begin{center}
\resizebox{13cm}{!}
{\includegraphics{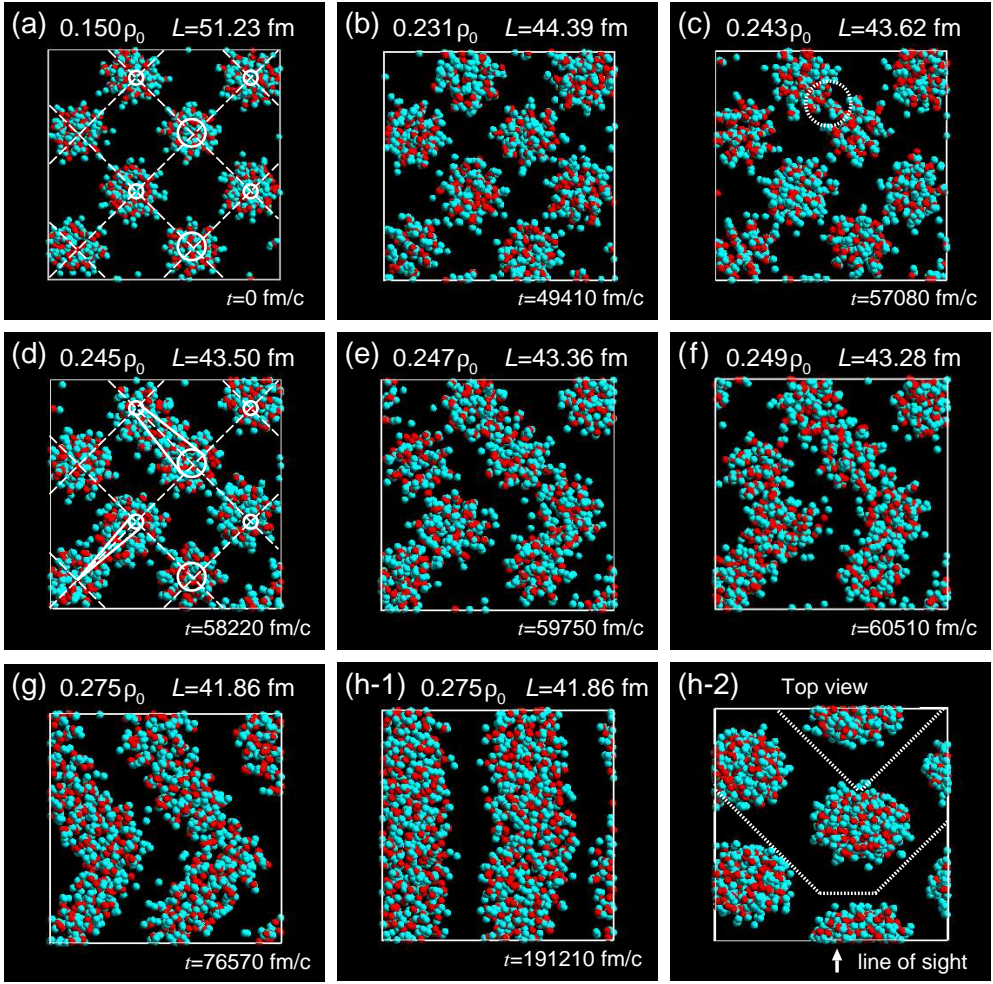}}
\caption{\label{fig_snapshot}
Snapshots of the formation process of the pasta phase with rod-like nuclei from a bcc lattice of spherical nuclei by compression of matter.
The red particles show protons and the green ones show neutrons.
In panels (a)-(g) and (h-1), nucleons in a limited region [surrounded by the dotted lines in panel (h-2)] are shown for visibility.
The vertices of the dashed lines in panels (a) and (d) show the equilibrium positions of nuclei in the bcc lattice and their positions in the direction of the line of sight are indicated by the size of the circles: vertices with a large circle, with a small circle, and those without a circle are in the first, second, and third lattice plane, respectively.
%The dotted circle in panel (c) shows the first pair of
%nuclei start to touch.
The solid lines in panel (d) represent the direction of the two elongated nuclei: they take zigzag configuration.  
The box sizes are rescaled to be equal in the figures.
This figure is taken from Ref.\ \citen{QMD-wata-formation}.
}
\end{center}
\end{figure}

In Fig.\ \ref{fig_snapshot}, we show the snapshots of our simulation, which capture the formation of the pasta phase in adiabatic compression.
Here we use the QMD Hamiltonian of Ref.\ \citen{QMD-maru} with the standard medium EOS parameter set as in the previous section.
The proton fraction $x$ and the total number of particles $N$ are $x\simeq 0.39$ and $N=3328$ (with 1312 protons and 2016 neutrons) in this simulation.
Starting from an initial condition at $\rho=0.15 \rho_0$ and $T=0.25$ MeV [Fig.\ \ref{fig_snapshot}(a)], we increase the density by changing the box size $L_{\rm cell}$ slowly at a rate $\lesssim \mathcal{O}(10^{-6})\ \rho_0/($fm$/c)$, which yields the time scale of $\gtrsim 10^{5}$ fm$/c$ to reach the typical density region of the phase with rod-like nuclei.%
\footnote{
While this time scale is, of course, much smaller than the actual time scale of the collapse, it is much larger than that of the change of the nuclear shape (e.g., $\sim 1000$ fm$/c$ for the nuclear fission) and thus the dynamics observed in the simulation should be governed by the intrinsic physical properties of the system, not by the density change applied externally.}
At $\rho\simeq 0.243 \rho_0$ [Fig.~\ref{fig_snapshot}(c)], the first pair of two nearest-neighbor nuclei start to touch and fuse (dotted circle), and then form an elongated nucleus.  
Then, multiple pairs of nuclei fuse and become such elongated nuclei in a way that they are aligned in a zigzag configuration [Fig.~\ref{fig_snapshot}(d)].
These elongated nuclei further stick together [see Figs.~\ref{fig_snapshot}(e) and (f)], and all the nuclei fuse to form rod-like nuclei as shown in Fig.~\ref{fig_snapshot}(g).  
Finally, we obtain an almost perfect triangular lattice of rod-like nuclei after relaxation [Figs.~\ref{fig_snapshot}(h-1) and (h-2)].

Note that, until the nuclei touch and fuse, they keep the spherical shape [see Fig.\ \ref{fig_snapshot}(c)].
This shows that the pasta phase is formed without undergoing the fission instability.
It is also remarkable that, in the middle of the transition process, the elongated nuclei made of a pair of spherical nuclei take a zigzag configuration and then they further connect to form wavy rod-like nuclei.
This process is very different from the above-mentioned conjectured scenario based on the fission instability.

\subsection{Expansion of nuclear matter\label{qmd-expand}}

In this section, we present another example of a MD simulation of matter, i.e., expanding nuclear matter \cite{Chikazumi00,Chikazumi01}.
Multifragmentation is one of the topics of a long-standing interest in heavy-ion collision physics. 
Not only does its mechanism itself attract our interest but also it is a good test bed for the EOS.
In particular, fragment mass spectra have been discussed in many works. 
%
%
%%The fragmentation mechanism is dependent on the region where they take place, i.e.\ the participant region or the spectator region.
%%In the participant region where colliding two nuclei overlap, the fragmentation is affected by the radial flow which makes one unable to assume a thermal equilibrium. 
%
The fragmentation mechanism is different between the participant region where colliding two nuclei overlap, and the spectator region which surrounds the participant region. 
In the participant region, the fragmentation is affected by a radial flow, and thus this region is not in thermal equilibrium.
%
%
%Empirically, the mass spectra with a large radial flow yield exponential shape \cite{Petrovici,Reisdorf}. 
Empirically, a large radial flow yields an exponential mass spectra \cite{Petrovici,Reisdorf}. 
On the other hand, 
%%in spectator region which does not collide with the other nucleus, statistical models\footnote{Stochastic decay of excited nuclei by particle evaporation is calculated.} are effective since the collective dynamics are not important. 
for the spectator region, statistical models are effective since the collective dynamics is not important.
Fisher's droplet model, which is one of the most famous statistical models, predict a power law of fragment mass spectra.

There are two major scenarios for fragment formation. 
%% which are widely accepted.
One is as follows: by the collision of two nuclei, a hot and dense zone is made.
It expands due to the high temperature and high densities. 
At this point, the system is considered as a gas.
%%As it becomes dilute and the temperature decreases, it crosses the boundary of gas and liquid.
As the system becomes dilute, the temperature decreases and finally crosses the liquid-gas transition point.
Then the system experiences the spinodal instability and fragmentation occurs.
The other scenario is that the expanding region behaves like a solid.
Then the fragmentation occurs by the formation of cracks.
%%So we have above two different scenarios for expanding systems.

Many studies have been performed so far for collision analysis \cite{Aichelin,Bonasera,Chomaz,Mishustin,Gilkes,Mastinu,Gregoire,Belkacem,Strachan,Colonna} which include both dynamics and statistics. 
However, the dynamics are complicated due to the finite size effect.
In order to  simplify the problem and to get more direct insights into the fragmentation mechanism, we perform QMD simulations of infinite matter by imposing special periodic boundary conditions with an isotropic expansion \cite{Chikazumi00,Chikazumi01}.
%%At the same time, by using the QMD model with the presence of the uniform expansion of the system, we take into account the essential dynamics of the system.
%
Prior to our study, instability of nuclear matter against multifragmentation has been studied \cite{Colonna2} and also a simulation study of the infinite system based on a dynamical model has been tried by other authors \cite{Finocchiaro}. 
%These studies, however, have not taken into account the expansion of the system.
These studies, however, do not take into account the expansion of the system.
%%Here we simulate expanding nuclear matter by means of the QMD model with periodic boundary conditions. 

\subsubsection{Model}

In order to simulate expanding nuclear matter, we employ a special periodic boundary condition. 
%The simulation of an expanding periodic system has been proposed and applied to two-dimensional (2D) systems in condensed matter physics \cite{Dorso,Brad}. 
A method to simulate two-dimensional (2D) expanding periodic systems has been proposed in condensed matter physics \cite{Dorso,Brad}. 
We extend this method for 3D systems and apply it to nuclear fragmentation \cite{Chikazumi00,Chikazumi01}.

Hamiltonian used here consists of the following effective interaction terms,
%to describe properties of nuclear matter and nuclei as 
%
\begin{eqnarray}
H
&=&
\sum_{i=1}\frac{{\bf P}_i^2}{2M}
+
V_{\rm nucl}+V_{\rm surf}+V_{\rm Pauli},
%\nonumber\\
\\
V_{\rm nucl}
&\equiv&
\frac{\alpha}{2\rho_0}
\sum_{i=1}
\langle \rho_i \rangle
+
\frac{\beta}{(1+\gamma)\rho_0^\gamma}
\sum_{i=1}
\langle \tilde\rho_i \rangle^{\gamma}
%\nonumber\\
+
\frac{C_{\rm s0}}{2\rho_0}
\sum_{i,j\neq i}
(1-2|c_i-c_j|)
\rho_{ij}
\nonumber\\
&+&
\frac{C_{{\rm ex}(1)}}{2\rho_0}
\sum_{i,j\neq i}
\frac{1}
     { 1
      +\left(
              \frac{|{\bf P}_i-{\bf P}_j|}{\mu_1}
       \right)^2
     }
\rho_{ij}
%\nonumber\\
+
\frac{C_{{\rm ex}(2)}}{2\rho_0}
\sum_{i,j\neq i}
\frac{1}
     { 1
      +\left(
              \frac{|{\bf P}_i-{\bf P}_j|}{\mu_2}
       \right)^2
     }
\rho_{ij},
%\nonumber\\
\\
V_{\rm surf}
&\equiv&
\frac{V_{\rm SF}}
     {2\rho_0^{5/3}}
\sum_{i,j\neq i}
%\int\int d^3r\;d^3r'
\int d^3r
\nabla \rho_i({\bf r})\cdot
\nabla \rho_j({\bf r}),
%\nonumber\\
\\
V_{\rm Pauli}
&\equiv&
\frac{C_{\rm P}          }
     {2(q_0p_0)^3}
\sum_{i,j\neq i}
%\nonumber\\
~
\exp\left[
%\left(
      -\frac{|{\bf R}_i-{\bf R}_j|^2}
            {2q_0^2                  }
      -\frac{|{\bf P}_i-{\bf P}_j|^2}
            {2p_0^2                  }
%\right)
\right]
\delta_{\tau_i,\tau_j}
\delta_{\sigma_i,\sigma_j},
%\nonumber\\
\end{eqnarray}
where $\langle\rho_i\rangle\equiv\sum_{j(\neq i)}\rho_{ij}$ are defined in Eqs.\ (\ref{rhoij}) and (\ref{rhoijtilde}).
Although this Hamiltonian is almost the same as that used in the previous sections (introduced in Sec.\ \ref{secEffectiveInt}), a surface term $V_{\rm surf}$ is newly added and parameters are readjusted as shown in Table \ref{table1c}. 
By this improvement of the effective interaction, experimental data of the radii and the  binding energies of nuclei are reproduced better. 
Coulomb potential and a two-body collision term are not included for simplicity. 

\begin{table}
\begin{center}
\caption{\label{table1c}
Effective interaction parameter set ($K$=280 MeV).}
%\begin{tabular}{|cc|cc|c|cc|}
\begin{tabular}{ cc cc c cc }
\hline
& $\alpha$ (MeV) &$-121.9$&
& $\beta$ (MeV) &$197.3$&\\
& $\gamma$ &$4/3$&
& $C_{\rm s0}$ (MeV) &$25.0$&\\
& $C_{{\rm ex}(1)}$ (MeV) &$-258.5$&
& $C_{{\rm ex}(2)}$ (MeV) &$375.6$&\\
& $\mu_1$ (MeV) &$2.35$&
& $\mu_2$ (MeV) &$0.4$&\\
& $V_{\rm SF}$ (MeV) &$20.68$&
& $C_{\rm P}$ (MeV) &$115.0$&\\
& $p_0$ (MeV) &$120.0$&
& $q_0$ (fm) &$2.5$&\\
%& $L$ (fm$^2$) &$1.95$&
& $\lambda$ (fm) &$1.40$&
&              &      &\\          
\hline
\end{tabular}
\end{center}
\end{table}

\begin{figure}
\begin{center}
\includegraphics[width=0.5\textwidth]{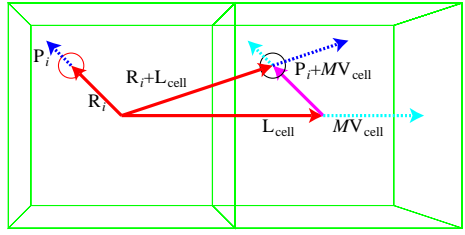}
\caption{
	Schematic explanation of the periodic boundary condition with uniform expansions.
	The left box is the primitive cell and the right one is a replica cell located at ${\bf L}_{\rm cell}$.
	The cell velocity is ${\bf V}_{\rm cell}\equiv\dot{\bf L}_{\rm cell}$.
This figure is taken from Ref.\ \citen{Chikazumi01}.
\label{fig4c}
}
\end{center}
\end{figure}

Numerical calculations are done in the primitive cubic cell and 26 replica cells surrounding the primitive cell. 
With the standard periodic boundary condition, we first prepare a set of equilibrated initial conditions of symmetric nuclear matter at $\rho_0$ for various values of the temperature.
In this model, we set the nuclear saturation density $\rho_0=0.16$ fm$^{-3}$.
According to the Metropolis sampling procedure, 1000 samples are created for each temperature. 
Next, we give each nucleon an additional collective momentum proportional to its position vector.
%The same motion is given to each cell boundary to achieve a consistent homogeneous expansion as a whole system.
The same motion is given to each cell boundary in order that the whole system undergoes a homogeneous expansion.
The collective momentum ${\bf P}_{\rm coll}$ of a particle $i$ in the replica cell is also proportional to the position vector ${\bf R}_i+{\bf L}_{\rm cell}$ of the particle [see Fig.\ \ref{fig4c}].
%
%The speed of expansion is characterized by a radial flow velocity parameter $h$, which is analogous to a cosmological term, the Hubble constant. 
The speed of expansion is characterized by a radial flow velocity parameter $h$, which is analogous to the Hubble constant in cosmology. 
The collective momentum added to a particle at a position ${\bf R}$ is as follows:
\begin{eqnarray}
	{\bf P}_{\rm coll}({\bf R})\equiv h\frac{{\bf R}}{\rho_0^{-1/3}}P_{\rm F}, \label{eqPcoll}
\end{eqnarray}
where $P_{\rm F}$ is the Fermi momentum at saturation density $\rho_0$. 
Each sample with a given temperature will be expanded with radial flow velocity parameter $h$. 
Note that when a nucleon goes out of a boundary of the primitive cell, its image particle comes in from the opposite boundary of the cell.
What is different from the normal periodic boundary conditions is that the momentum of the image particle is also modified by the collective momentum ${\bf P}_{\rm coll}({\bf L}_{\rm cell})$ proportional to the cell size.

In our setup, this collective motion does not stop.
Moreover, it does not change at all.
This is because the collective force per unit cell which intends to change the collective momentum is finite but the moment of inertia per unit cell diverges for the infinite system.
In other words, while the potential energy per cell is finite, the kinetic energy per cell is infinite due to the contributions of infinitely distant cells with diverging velocities.
%One may think that this statement contradicts the reality of stellar compression or expansion.
%However, the existence of the surface plays a crucial role in the acceleration of the collective motion.
%We cannot avoid this constant collective velocity if matter with periodic boundary conditions is used. 
We cannot avoid this constant collective velocity when we consider expansion or contraction of infinite systems with periodic boundary conditions.% 
% which represents an infinite system is used to mimic the finite but large-scale matter.
%\footnote{Note that in the calculation of compressing nuclear matter (Sec.\ \ref{sect_qmdsn}), the collective motion (compression) was not provided according to this rule but by arbitrary manner.
%This is because the compression of the system in the simulation was very slow compared to the motion of nucleons.}
\footnote{In the simulation of adiabatic compression of nuclear matter (Sec.\ \ref{sect_qmdsn}), we have rescaled the particle positions at the compression rate instead of using the initial collective momentum explained in this section. 
This is valid because the compression of the system in that case is very slow compared to the motion of nucleons.}

\begin{figure}
\begin{center}
\includegraphics[width=0.5\textwidth]{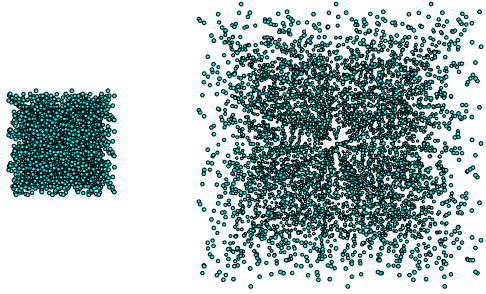}
\caption{
Snapshots of nucleon distribution in the primitive cell in the initial state at $\rho_0$ (left panel) and the final state at $0.05\rho_0$ (right panel).
This figure is adapted from Ref.\ \citen{Chikazumi01}.
\label{fig3c}
}
\end{center}
\end{figure}

We calculate the time evolution  until the average density reaches $0.05\rho_0$, below which we can identify fragments.
Figure \ref{fig3c} shows an example of our simulation; the initial state at $\rho_0$ and $T=30$ MeV (left panel) and the final state at $0.05\rho_0$ (right panel).
Each fragment is determined by the clustering algorithm:
When the distance between a particle and a fragment is smaller than 3 fm, we identify that this particle belongs to the fragment.
We perform 1000 events for each $h$ and $T$ to get fragment mass spectra. 

\subsubsection{Results}

Before discussing how fragment mass spectrum depends on $h$ and $T$,
we study how the pressure of the system is related to the dynamical expansion.
The instantaneous pressure is calculated on the basis of the virial theorem as follows \cite{Hoover}:
\begin{eqnarray}
P = \frac{2}{3N} \rho \sum_{i} \frac{{\bf v}_i \cdot {\bf P}_i^{\rm thermal}} {2} + \frac{1}{6N} \rho \sum_{i} \sum_{j} ({\bf R}_i-{\bf R}_j) \cdot {\bf F}_{ij},
           \label{eqPress}
\end{eqnarray}
where ${\bf v}_i\equiv\dot {\bf R}_i$ is the velocity of particle $i$ and ${\bf F}_{ij}$ is the force acting between particles $i$ and $j$. 
%Although the usual definition of pressure using the virial theorem requires time averaging, it is also possible to define the instantaneous pressure as given by Eq.\ (\ref{eqPress})\cite{Hoover} and we use this definition in our calculation. 
Note that the momentum includes two components: One is the momentum of the thermal motion, and the other is that of the collective motion characterized by $h$. 
To calculate the  pressure, we subtract the momentum of the collective component of Eq.\ (\ref{eqPcoll}) from ${\bf P}_i$, i.e., ${\bf P}^{\rm thermal}_i\equiv {\bf P}_i-{\bf P}^{\rm coll}_i$ in Eq.\ (\ref{eqPress}).  

\begin{figure}
\begin{center}
\includegraphics[width=0.65\textwidth]{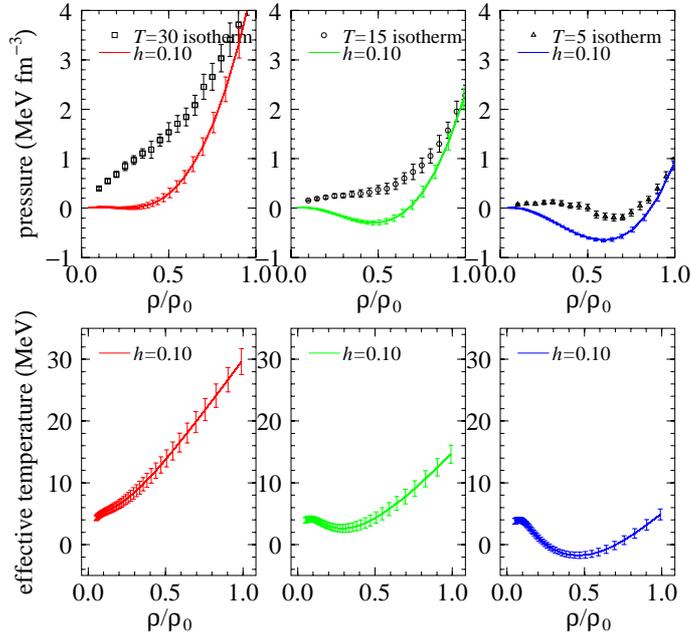}
\caption{
Upper panels: Density dependence of the pressure.
From the left, the initial temperature $T_{\rm init}=30$, $15$, and $5$ MeV.
Dots show isotherms which are calculated at each fixed density by the Metropolis sampling method. 
Lines show adiabatic cases that nuclear matter expands 
from saturation density $\rho_0$ to $0.05\rho_0$. 
Lower panels: Density dependence of the effective temperature.
This figure is taken from Ref.\ \citen{Chikazumi01}.
\label{fig1}
}
\end{center}
\end{figure}
\begin{figure}
\begin{center}
\includegraphics[width=0.7\textwidth]{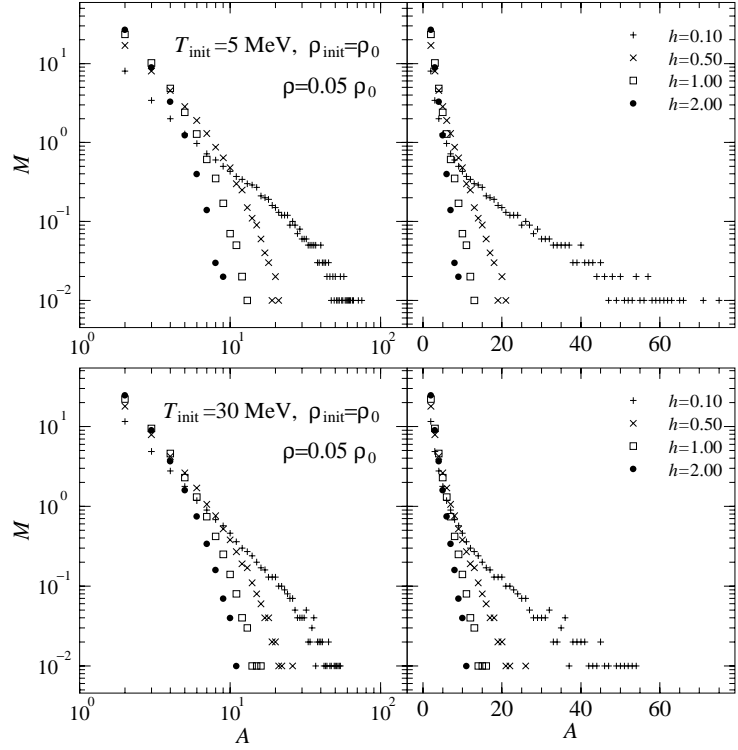}
\caption{
The fragment mass spectra at $\rho=0.05\rho_0$.
The fragment multiplicity $M$ is counted in the primitive cell.
The expansion begins at saturation density $\rho_0$ with the initial configuration created by Metropolis sampling method. 
Upper figure shows the results for initial temperature $5$ MeV and lower one for $30$ MeV. 
The radial flow velocities corresponding to $h=0.10$, $0.50$, $1.00$, and $2.00$ are shown. 
Both the right and the left show the same data with different scale of abscissa.  
This figure is taken from Ref.\ \citen{Chikazumi00}.
\label{fig2}
}
\end{center}
\end{figure}

Figure \ref{fig1} shows the relation between density and pressure. 
Bold lines are isothermal ($T=5$, $15$, and $30$ MeV) cases calculated by the Metropolis sampling method at each density.  
Thin lines are the trajectories along which the system expands.
The instantaneous pressure of the expanded matter is smaller than that of the equilibrium state. 
%%The pressures of these dynamical expansion are below those of isotherms except for their starting points at $\rho_0$. 
%
One reason of this reduction is that the expansion of matter proceeds adiabatically, i.e.,  without an exchange of heat and with a positive work to the environment. 
Therefore, the system is cooled down as it expands.%
\footnote{
In the case of low initial temperatures, there appears a region of negative effective temperatures (see the lower right panel of Fig.\ \ref{fig1}). 
This means that some particles have velocity and momentum of opposite directions, which is caused by the momentum-dependent terms of the potential.}
Another reason is that the expansion is a dynamical and non-equilibrium process.
%If the system were static, the region of mechanical instability with a negative gradient of the pressure against the density would disappear by a formation of clusters.
If the system were static, the region of mechanical instability with a negative gradient of the pressure against the density would be avoided by a formation of clusters.%
\footnote{
One may notice that the pressure in the upper right panel of Fig.\ \ref{fig1} still has a negative gradient.
However, with inclusion of electron contribution, the total pressure shows monotonically increasing dependence on the density.
Similar situation is reported in Ref.\ \citen{Maruyama12}.
}
For the expanding case, on the other hand, the system cannot reach the stable state with a higher pressure.  
This is the same tendency observed experimentally in Ref.\ \citen{Shlomo}.     
%% If we increase the radial flow up to $h=2.0$, the relation between density and pressure does not depend on the flow velocity within the accuracy of 0.1 MeV fm$^{-3}$.            
%For large radial flow velocities of $h>2.0$, the relation between the density and the pressure does not depend on the flow velocity within the accuracy of $0.1$ MeV fm$^{-3}$.            

The fragment mass distributions   are very sensitive to the radial flow velocity.
Figure \ref{fig2} shows the fragment mass distributions 
%in the case of dynamical expansion calculated at $\rho=0.05\rho_0$.
obtained at $\rho=0.05\rho_0$ after the dynamical expansion.
Both the right and the left panels show the same data but with different scales of abscissa.
As the radial flow velocity increases, the fragment mass distribution changes from the power law, i.e., a straight line in the double logarithmic scale, to the exponential decay, i.e., a straight line in the linear-logarithmic scale.
%The increase of its slope with the increase of $h$ should also be noted.
%
It is also noted that the fragment mass distribution decreases more rapidly with increasing $h$.
This feature is consistent with experimental data of heavy-ion collisions at small impact parameters where a large radial flow is observed \cite{Reisdorf}.
% As we can see from the figure, the mass spectra show only small dependence on the initial temperature.
%

In the fast expansion limit, the effect of the interaction between particles is expected to be small and the fragmentation is determined solely by the initial position of particles.
The fragment mass distribution in this limit shows an exponential form \cite{Brad} and the change of the slope according to the expansion velocity is also consistent with the consideration given in Ref.\ \citen{Mastinu}.

In the case of slow expansion ($h=0.1$), the fragment mass distributions for $T_{\rm init}=5$ and $30$ MeV agree very well to the Fisher's power law.
One may conclude that these fragment mass distributions indicate an evidence of the liquid-gas phase transition of nuclear matter.
However, these results do not always support this scenario.
The reason is as follows:
%%One important premise of the Fisher's model is that the temperature is above the critical temperature which is estimated to be $\simeq 8$ MeV in our QMD.
%%%%%Although the value of this critical temperature is not clear, it must be much higher than $5$ MeV. 
%%In our case with an initial temperature of $5$ MeV, where the mass distribution shows a power law, the freeze-out temperature should be lower than $5$ MeV. 
%%Therefore, the fragment mass distribution with the power low does not necessarily phase transition.
%%Instead, a breakup of a solid \cite{Oddershede,Timar} by cracking is more plausible in the case of lower temperatures. 
%
One important premise of the Fisher's model is that the temperature is above the critical temperature, which is about $8$ MeV in the present model \cite{Chikazumi01}.
On the other hand, in our case with the initial temperature of $5$ MeV, where the mass distribution shows a power law, the freeze-out temperature should be lower than $5$ MeV.
Therefore, the fragment mass distribution with the power low is not necessarily accompanied by the liquid-gas phase transition.
Instead, a breakup of a solid \cite{Oddershede,Timar} by cracking is more plausible in the case of lower temperatures.

\section{Summary}

In this article, we have reviewed the molecular dynamics approaches to nucleon many-body systems, focusing on the quantum molecular dynamics (QMD) model.
This method was originally developed for the study of heavy-ion collisions to describe multifragmentation which the time-dependent Hartree-Fock (TDHF) theory failed to explain.
With the help of the Pauli potential to take account of the Pauli principle, QMD has been applied to dense nuclear matter in addition to heavy-ion collisions.

A great advantage of MD approaches is that we can study the dynamical process of nucleon many-body systems without any assumptions on the nuclear structure. 
QMD, in particular, is very suitable to study inhomogeneous nuclear matter in the mesoscopic to macroscopic scales because QMD makes it possible to simulate large systems with many nucleons.
To show the power of this method, we have presented some of our works using QMD.

First, we have explained our QMD model in Sec.~\ref{sec-formulation}.
We have seen that this model is designed to give the correct saturation properties and reasonable equations of state (EOS) of nuclear matter, and give a good agreement of the binding energies of light nuclei with $A\lesssim 8$ including alpha particle and also heavy nuclei with $A\gtrsim 40$.
These points are important for the reliability of the predictions by this model, which we have presented in the remaining part of this article.

 From Sec.~\ref{QMDmatter-maru} to \ref{sect_qmdsn}, we have shown a series of our studies about the structure of nuclear matter at subsaturation densities. 
It has been predicted that nuclear ``pasta'' phases, states of matter with nuclei of rod-like and slab-like shapes, can be the ground state of matter in this density region.
Using QMD, we have shown that the pasta phases can actually be formed by cooling hot uniform nuclear matter and by compressing a bcc lattice of spherical nuclei.  
These results strongly suggest that the pasta phases are formed in the cooling process of hot neutron star crusts and by the compression of matter in the collapse of supernova cores.

These results have important implications on the mechanical strength of the neutron star crust,
the cooling process of hot protoneutron stars, the mechanism of glitches, etc.
Making an EOS table for core collapse simulations, taking into account the existence of the pasta phases including their effects on the neutrino opacity, is an important direction.

In Sec.~\ref{qmd-expand}, we have discussed fragment formation in expanding nuclear matter. 
We have developed a method to describe isotropically expanding matter using the periodic boundary conditions.  
Using this method together with our QMD model, we have calculated the fragment mass distribution and have found that it shows an exponential decay for rapid expansion and a power-law decay for slow expansion.
%Our analysis suggests that multifragmentation occurs by the formation of cracks in the solid-like expanding region rather than the scenario based on the liquid-gas phase transition.
Our analysis suggests that multifragmentation at lower temperatures occurs not by the liquid-gas phase transition but by some other mechanisms, e.g., the formation of cracks in the solid-like expanding region.

%% The MD simulations are getting more popular and important in material science and technology.
%% The constituent particles are not limited to simple particles but can be polarized or can have complicated structures such as atoms and molecules. 
%% The types of systems also have a large variety; microcanonical, canonical, and grandcanonical ensembles, systems with energy dissipation, systems with expansion, compression, and distortion, etc.
%% In nuclear physics, the target and the method of simulation, including MD, should be full of interesting possibilities.

Molecular dynamics simulations have been playing important roles in nuclear physics: both in the study of the nuclear structure and reaction. 
They can successfully describe the structure of nuclei and statistical properties of heavy-ion collisions taking account of many-body correlations and fluctuations.
In addition to nucleon many-body systems, MD simulations are used also in QCD studies: UrQMD is now commonly used to analyze heavy-ion collisions with a quark-gluon plasma \cite{refUrQMD}, and some dynamical properties of quark matter have been studied by a MD approach\cite{qmmd,akimura}.
One of the most important and challenging directions is to incorporate the wave nature of the quantum mechanics in the MD approach, which is based on the particle picture.
A new framework beyond QMD and FMD/AMD in this direction is highly awaited \cite{AMDV}.
By such a breakthrough, MD should be a promising approach also for studying the dynamics of fission and fusion, 
which is a long-standing mportant problem in nuclear physics.

\section*{Acknowledgment}
We are grateful to 
S.~Chikazumi, C. O. Dorso, T. Ebisuzaki, K. Iida, A.~Iwamoto, 
K.~Niita, K. Oyamatsu, C. J. Pethick, K. Sato, H. Sonoda, T. Takiwaki, T. Tatsumi, and K. Yasuoka 
for collaborations and fruitful discussions.

This work was supported in part by the Max Planck Society, the Korean Ministry of Education, Science and Technology, Gyeongsangbuk-Do, Pohang City, for the support of the JRG at APCTP and by Basic Science Research Program through the National
Research Foundation of Korea (NRF) funded by the Ministry of Education, Science and Technology (No. 2012008028).
Calculations were performed on RIKEN Super Combined Cluster System with MDGRAPE-2 and -3.

\end{document}